\documentclass[12pt]{iopart}

\usepackage[dvips]{graphicx}
\usepackage{amsmath,amssymb,amsthm,mathrsfs,amsfonts,dsfont,amscd,keyval}
\usepackage{mathtools,mathrsfs}
\usepackage{yfonts} 
\usepackage{ytableau}
\usepackage{subfigure, epsfig}
\usepackage{braket}
\usepackage{bm}
\usepackage{fancyhdr}
\usepackage{enumerate}
\usepackage{color}
\usepackage{hyperref}
\usepackage{tabularx}
\usepackage{times}
\hypersetup{colorlinks=true, linkcolor=blue, citecolor=blue, urlcolor=black}
\usepackage{multirow}
\usepackage{float}
\usepackage{tensor}
\usepackage{multirow}
\usepackage[english]{babel}
\usepackage[square,numbers]{natbib}
\newcommand{\overbar}[1]{\mkern 1.5mu\overline{\mkern-1.5mu#1\mkern-1.5mu}\mkern 1.5mu}

\theoremstyle{definition}
\newtheorem{definition}{Definition}[section]

\theoremstyle{plain}
\newtheorem{theorem}[definition]{Theorem}

\theoremstyle{remark}

\newcommand{\comments}[1]{}
\newcommand{\ketbra}[2]{\vert #1 \rangle \langle #2 \vert}

\newcommand{\dbar}{d\hspace*{-0.08em}\bar{}\hspace*{0.1em}}

\begin{document}

\title{A comprehensive review of Quantum Machine Learning: from NISQ to Fault Tolerance}

\author{Yunfei Wang}

\address{Martin A. Fisher School of Physics, Brandeis University, Waltham MA 02453, USA}

\ead{yunfeiwang0214@gmail.com}

\author{Junyu Liu}

\address{Chicago Quantum Exchange, Chicago, IL 60637, USA}

\address{Department of Computer Science, The University of Chicago, Chicago, IL 60637, USA}

\address{Pritzker School of Molecular Engineering, The University of Chicago, Chicago, IL 60637, USA}

\address{Kadanoff Center for Theoretical Physics, The University of Chicago, Chicago, IL 60637, USA}

\ead{junyuliu@uchicago}

\vspace{10pt}

\begin{abstract}
Quantum machine learning, which involves running machine learning algorithms on quantum devices, has garnered significant attention in both academic and business circles. In this paper, we offer a comprehensive and unbiased review of the various concepts that have emerged in the field of quantum machine learning. This includes techniques used in Noisy Intermediate-Scale Quantum (NISQ) technologies and approaches for algorithms compatible with fault-tolerant quantum computing hardware. Our review covers fundamental concepts, algorithms, and the statistical learning theory pertinent to quantum machine learning.
\end{abstract}

\maketitle

\tableofcontents

\section{Introduction}

Quantum machine learning represents a highly promising realm in contemporary physics and computer science research, with far-reaching implications spanning quantum chemistry \cite{peruzzo2014variational}, artificial intelligence \cite{liu2023towards}, and even high-energy physics \cite{andreassen2019junipr}. Nevertheless, it remains in its nascent stages of development. This is evident from the absence of a precise definition for quantum machine learning. Some describe it as the convergence of quantum computing and machine learning, wherein machine learning algorithms are executed on quantum devices. In simpler terms, it can be thought of as the quantum counterpart to classical machine learning.

In recent times, artificial intelligence, exemplified by technologies like \texttt{ChatGPT}, has become an integral part of everyday life. It's entirely plausible that, in the future, we will harness artificial intellegence in an even wider array of applications, including medical diagnostics, education, and aiding scientific research. Much of artificial intellegence's success hinges on machine learning, a departure from traditional computer programs that entail crafting explicit instructions to directly solve problems \cite{roberts2022principles}. Machine learning models, on the other hand, are trained on real-world data stored in a dataset (often denoted as $\mathcal{D}$), acquiring the ability to tackle problems autonomously. This represents a departure, in a sense, from the conventional von Neumann model of digital computing.

Researchers might initially wonder why there's a need for quantum machine learning when its classical counterpart has demonstrated impressive performance. One of the primary rationales stems from the fact that classical machine learning relies heavily on linear algebra procedures \cite{biamonte2017quantum}. For instance, classical machine learning can address problems like identifying the optimal hyperplane to separate two data clusters by employing matrix inversion techniques. Quantum mechanics, at its core, is inherently grounded in linear algebra, where we construct a Hilbert space ($\mathcal{H}$) of a system described by an Hermitian operator known as the Hamiltonian ($\hat{H}$) \cite{Sakurai:2011zz}. Over time, it has become evident that quantum computing has the potential to dramatically enhance a computer's problem-solving capabilities in certain specific scenarios. To illustrate, consider the task of matrix inversion again. Classical computers typically require computational efforts with a complexity of $\mathcal{O}(N\log{N})$ to accomplish this, while quantum computers can leverage algorithms like the Harrow-Hassidim-Lloyd algorithm (HHL) algorithm \cite{harrow2009quantum}, which has a complexity of $\mathcal{O}((\log{N})^2)$, leading to significant speedup, in certain conditions like sparsity, fault-tolerance, small condition numbers, and fast interfaces between classical and quantum processors. Although there is no known explicit realizations of HHL algorithm at the large scale at the moment, it is still an exciting direction deserves further studies, especially when quantum devices are fast developing with more and more capabilities. 

Due to its vague definition, quantum machine learning encompasses a broader spectrum of topics. For instance, quantum (shadow) tomography \cite{aaronson2018shadow} has gained prominence, focusing on the characterization of a given quantum state by accumulating data from various measurements. This involves determining the minimum number of identical quantum state copies needed to extract sufficient information about the state's properties. Another facet involves machine learning for quantum physics, which entails employing (perhaps classical) machine learning tools to explore various aspects of quantum physics. Additionally, some developments in both quantum algorithms and quantum hardware are often encompassed within the broader umbrella of quantum machine learning.

Although these topics may not directly align with the narrow definition of quantum machine learning outlined earlier, they hold substantial promise for its future development. To illustrate, within the narrow sensed field of quantum machine learning, a significant challenge is known as the input/output problem. In particular, the output problem pertains to the task of comprehending the solution generated by a quantum algorithm, and this aligns closely with the domain of shadow tomography. And by harnessing the capabilities of classical machine learning to gain deeper insights into quantum physics, it might lead to advancements in quantum computing too. In this review, we primarily focus on quantum machine learning in its narrower sense, which pertains to execute quantum algorithms designed for machine learning purposes.

Presently, we find ourselves in what's referred to as (perhaps the end of) the \textit{noisy intermediate scale quantum} or NISQ era of quantum computing. Quantum computers are susceptible to background noise, which imposes limitations on our ability to construct quantum computers with sufficient depth for executing tasks demanding fast and precise computations. The quantum computers available today can only handle on the order of around 100 qubits, and they all exhibit noise, making it challenging to derive tangible benefits for our daily lives.

The solution to this predicament is known as quantum error correction (QEC) code \cite{Roffe_2019}. Think of QEC as a safeguard for quantum information. Typically, quantum information is lost once it's measured, which becomes especially likely in noisy environments. However, information protected by QEC can persist if it remains undamaged within certain limits. It's worth noting that all error correction codes have their constraints, implying that information will inevitably be lost if it's severely damaged. Nevertheless, error correction provides a protective buffer zone against such losses.

Our objective is to implement QEC codes across all quantum devices, ushering in an era of \textit{fault-tolerant quantum computing} (FTQC) in the future. This trajectory parallels the history of classical computing, where, before the invention of classical error correction, scaling up and running useful algorithms on classical computers was also a formidable challenge \cite{chang2006introduction}. Today, we can reliably operate classical computers everyday. Given the promising advancements in quantum error correction in recent times, our optimism about the future of quantum computing remains steadfast.

However, when it comes to machine learning, classical machine learning doesn't inherently reject noise \cite{liu2022noise}. The widely recognized learning algorithm, known as stochastic gradient descent, explicitly incorporates noise, and surprisingly, this noise addition actually enhances its performance. To grasp this concept, consider that noise can effectively steer us away from saddle points, offering an automatic mechanism for avoiding them. In a way, one can interpret this as machine learning's ability to withstand and even benefit from noise. Consequently, this insight suggests that running certain machine learning algorithms on current (NISQ) quantum devices could have some significance. This prospect promises to enrich our present-day experiences significantly, especially considering the challenges associated with constructing QEC, which could take a few years to become fully integrated. Prior to entering the era of fault-tolerant quantum computing, we'll have the opportunity to experiment with quantum devices and apply machine learning techniques, adding a vibrant dimension to our current scientific research.

In addition to discussing the current applications of quantum capabilities for machine learning, we should also let our imaginations soar. The era of fault-tolerant quantum computing (FTQC) is a foreseeable future, and it's crucial to look ahead at what challenges we can tackle with quantum machine learning on FTQC devices. One of the most renowned and valuable algorithms in this context is the Harrow-Hassidim-Lloyd (HHL) algorithm \cite{harrow2009quantum}. Additionally, there are other algorithms that can be deployed to address a range of problems, such as principal component analysis \cite{biamonte2017quantum}.

Beside the optimistic future that quantum machine learning has, there are also a number of controversial issues with the subject. For example, some might argue that the variational quantum algorithm will not work in some circumstance. People working on quantum landscapes theory observed the famous barren plateau phenomena which leads to a kind of no-go theorem \cite{cerezo2022challenges,liu2022laziness}. It amounts to situations that are hard to find minimum when we optimizes our objective function by gradient descent with randomized variational circuits.

Debates indeed exist regarding whether quantum algorithms can consistently deliver exponential speedups. Some argue that quantum speedup is only guaranteed when dealing with quantum information. When it comes to classical information, it's conceivable to design classical machine learning models capable of achieving comparable average prediction accuracy \cite{tang2019quantum,huang2021information}. In such cases, the computational complexity difference between classical and quantum approaches may, at worst, be a modest polynomial factor. This debate underscores the importance of carefully assessing the specific problem and context when considering the potential advantages of quantum algorithms. However, the landscape shifts when the objective is to attain a low worst-case prediction error. In this scenario, it becomes feasible to achieve an exponential divergence in computational complexities between classical and quantum approaches. This underscores the potential superiority of quantum algorithms when stringent requirements for worst-case accuracy are in play. This phenomenon can be viewed as a manifestation of classical effects casting their ``shadow'' in the quantum realm. It should come as no surprise to individuals well-versed in quantum theory that quantum predictions tend to align with classical theory when dealing with systems possessing a high degree of freedom. This approach can also be applied to tackle a critical issue in variational quantum algorithms known as the ``output problem," often referred to as the classical shadow.

Indeed, it is truly heartening and captivating to witness the evolution of quantum computing beyond celebrated algorithms like Shor's factoring algorithm \cite{Shor1994Algorithm} and Grover search \cite{ChairmanMiller1996Proceedings}. Algorithms like HHL \cite{harrow2009quantum} offer a versatile blueprint, shedding light on how quantum computers hold the potential to deliver significant acceleration for fundamental tasks like clustering, pattern-matching, and principal component analysis. However, as Scott Aaronson aptly noted in his paper \cite{aaronson2015read}, the pursuit of these speedups will still demand significant effort and ingenuity, as nature continues to present challenges that compel us to work diligently for these advancements.

We have structured our review as follows. In section \ref{NISQ}, we focus on the historical and ongoing developments in Quantum Machine Learning (QML) during the Noisy Intermediate-Scale Quantum (NISQ) era. Within this era, one of the central frameworks is the Variational Quantum Algorithm (VQA) \cite{bharti2021noisy, farhi2014quantum, peruzzo2014variational}, which we explore in detail in section \ref{VQA}. VQA comprises four key elements: the choice of an objective function (discussed in section \ref{ObFunction}), the employment of Parameterized Quantum Circuits (PQC) with adjustable parameters optimized by classical algorithms (covered in section \ref{PQC}), measurement strategies (explored in section \ref{Measurements}), and the classical optimizer responsible for minimizing the objective function (explained in section \ref{Optimization}). Subsequently, we provide insights into the construction of the Quantum Neural Tangent Kernel (QNTK) \cite{liu2022representation} in section \ref{QNTK}, offering a theoretical foundation for quantum neural networks and an understanding of stochastic gradient descent dynamics from first principles. Finally, we address the issue of barren plateaus in section \ref{BarrenPlateaus}, approaching this topic through the lens of quantum landscape theory \cite{cerezo2022challenges}. Additionally, we present an alternative formulation of \textit{laziness} \cite{liu2022laziness}, providing another perspective on the same problem. 

In section \ref{FTQC}, our focus shifts to quantum algorithms that have the potential for exponential speedup in the Fault-Tolerant Quantum Computing (FTQC) era. We begin by introducing Quantum Phase Estimation (QPE) \cite{chang2006introduction} in section \ref{QPE&qPCA}, where we also delve into the Quantum Principal Component Analysis (QPCA) program \cite{lloyd1996universal}. Subsequently, we present a counterpoint regarding the attainability of exponential speedup for programs such as the recommendation system in section \ref{Dequantization}. Moving on to section \ref{HHL}, we introduce the pivotal Harrow-Hassidim-Lloyd (HHL) algorithm \cite{harrow2009quantum}, which opens up possibilities for various new quantum machine learning algorithms. In section \ref{Carleman}, we illustrate a visionary perspective on the future of machine learning algorithms with the aid of Carleman linearization \cite{Liu2021Efficient} and the HHL algorithm. Finally, in section \ref{QRAM}, we introduce Quantum Random Access Memory (QRAM), focusing on theoretical designs \cite{Hann1}, practical implementation efforts \cite{hann2021practicality}, and its necessity in certain algorithms \cite{biamonte2017quantum}.

In section \ref{Statistical}, we delve into various topics that amalgamate quantum principles with statistical learning theory. Our primary focus is on shadow tomography \cite{aaronson2018shadow} in section \ref{Shadow}, where we explore the motivation behind shadow tomography and delve into the construction of the theorem. A pivotal subject inspired by shadow tomography, the classical shadow formalism \cite{Elben_2022}, is thoroughly discussed in section \ref{CAlgorithms}. Moving on to section \ref{MLClassicalShadow}, we examine the application of classical shadow as an efficient quantum-to-classical information converter \cite{huang2021information}. In section \ref{QuantumData}, we then shift our attention to the applications of Quantum Machine Learning (QML) in the study of quantum data and quantum simulators \cite{biamonte2017quantum}. 

It's essential to note that our review doesn't encompass various topics, including quantum machine learning algorithms whose advantages stem from sampling-related statements, the studies of quantum error corrections, quantum memory, and designs of quantum data centers.

\section{Noisy intermediate-scale quantum (NISQ) era} \label{NISQ}

The concept of the NISQ era, which stands for Noisy Intermediate-Scale Quantum, was introduced by John Preskill in 2018 \cite{preskill2018quantum}. It's essential to understand that NISQ is primarily a hardware-oriented definition and doesn't inherently imply a specific time frame. In NISQ devices, quantum circuits are implemented, where all gates adhere to a predefined graph $G$, with qubits represented by nodes in this graph. These gates typically operate on one or two qubits. Due to the noise associated with each gate operation, NISQ algorithms are naturally constrained to shallow circuit depths \cite{barak2021classical}.

On the other hand, ``near-term" algorithms are those designed for quantum devices expected to be available in the next few years, and this term doesn't explicitly refer to the absence of Quantum Error Correction (QEC). The notion of what constitutes the ``near-term" can be subjective because different researchers may have varying opinions on how many years can be considered ``near-term." Predicting experimental progress in quantum computing is always challenging, and such predictions can be influenced by human biases. Algorithms developed for near-term hardware may become unfeasible if hardware advancements do not align with the algorithm's experimental requirements \cite{bharti2021noisy}.

The majority of existing NISQ algorithms leverage quantum computers' capabilities through a hybrid quantum-classical approach. These algorithms involve offloading the classically challenging aspects of certain computations to a quantum computer while performing the remaining tasks on a suitably powerful classical device. They operate by iteratively adjusting the parameters of a parametrized quantum circuit, making them known as Variational Quantum Algorithms (VQA) (\cite{endo2021hybrid,cerezo2022variational}).

\subsection{Variational quantum algorithm (VQA)} \label{VQA}

The concept of Variational Quantum Algorithms (VQA) was initially introduced to address certain quantum chemistry challenges, notably through the well-known Variational Quantum Eigensolver (VQE) \cite{mcclean2016theory,peruzzo2014variational,wecker2015progress}. Additionally, VQA includes the Quantum Approximate Optimization Algorithm (QAOA) \cite{farhi2014quantum}, which is designed for solving combinatorial optimization problems. Interestingly, reference \cite{Amaro_2022} shows that VQE combined with filtering operators can be more efficient and accurate than standard VQE and QAOA in the context of combinatorial optimization.

\begin{figure}[!ht]
\centering
\includegraphics[width=0.9\textwidth]{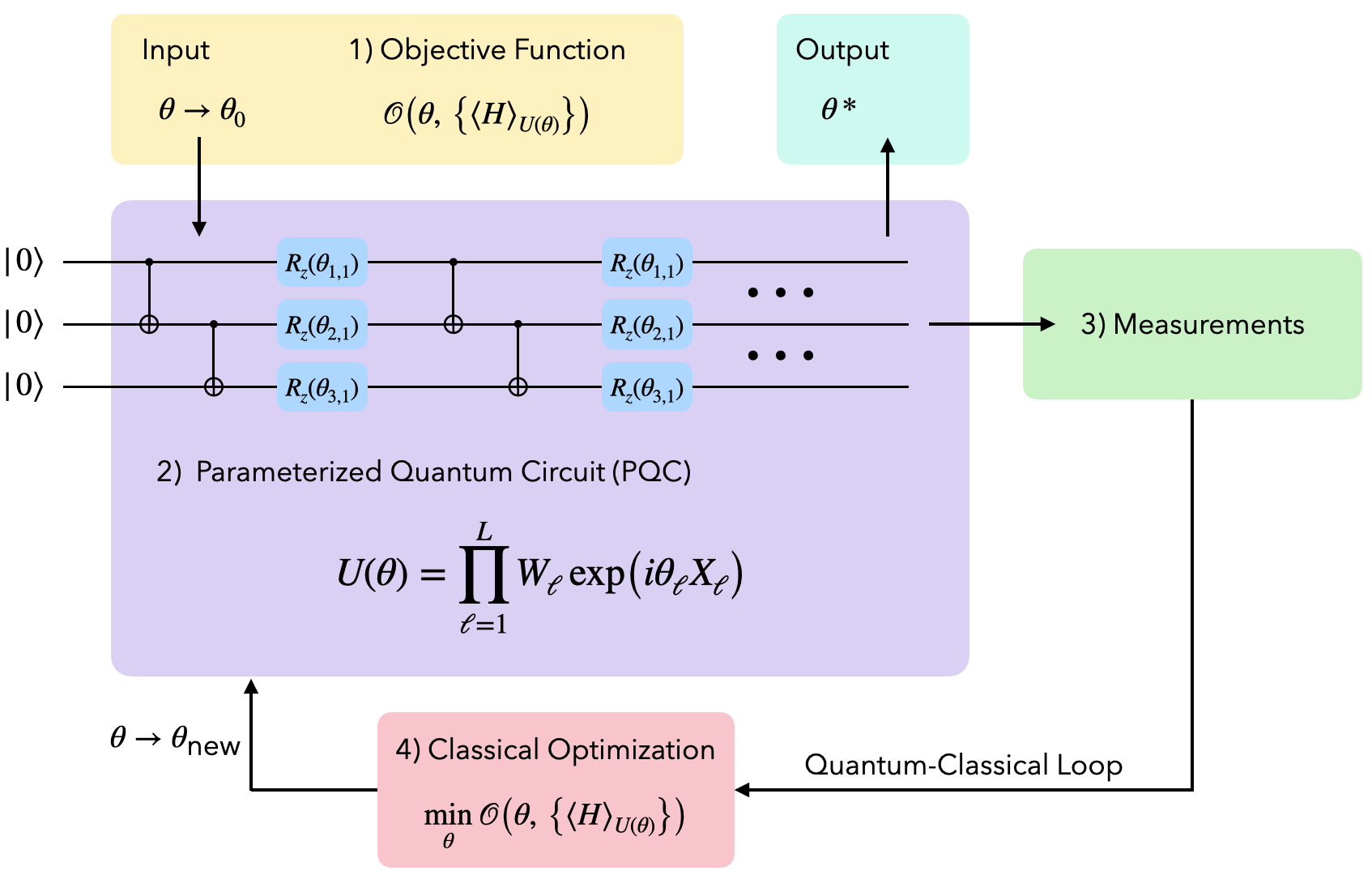}
\caption{VQA in 4 parts.}
\label{VQA4Parts}
\centering
\end{figure}

It's crucial to emphasize that, as of now, there is no established quantum advantage for Variational Quantum Algorithms (VQA) employing NISQ devices. As previously outlined, VQA can be divided into four primary components: 1) the objective function; 2) the parameterized quantum circuit (PQC); 3) the measurement scheme; and 4) the classical optimizer. This breakdown is visually represented in FIG.~\ref{VQA4Parts}. 

\subsubsection{Objective function} \label{ObFunction}

The Hamiltonian, a Hermitian operator, plays a pivotal role in governing the time evolution of a physical system through Schrödinger equation. Its expectation value provides the energy associated with a quantum state, making it a common target for minimization in VQA, often used to find the ground state of a system. However, when we aim to solve problems unrelated to real physical systems, we can construct a Hamiltonian that encodes the information we seek to extract.

In the context of VQA, the specific choice of the Hamiltonian is not particularly significant. One can select any Hermitian operator and designate it as the Hamiltonian, then attempt to find the ground state. However, an essential property that the operator must possess is the ability to decompose into several operators that can be measured by the quantum processor. We will explore this concept in more detail in section \ref{Measurements}. 

Constructing the objective function $\mathcal{O}$, it is natural to employ the expectation value of the Hamiltonian:

\begin{equation} \label{HamExpValue}
    \langle H \rangle_{U(\theta)} \equiv \bra{0} U^{\dagger}(\theta) H U(\theta) \ket{0}~.
\end{equation}

It's important to note that we initialize our qubits in the state $\ket{0}$ and subsequently evolve them using the parameterized quantum circuit (PQC). As shown in Eqn.~\ref{HamExpValue}, the objective function depends on the unitary time evolution operator, making it a function of classical tunable parameters denoted as $\theta$. In some situations, it can be advantageous to utilize objective functions other than the Hamiltonian expectation value. Consequently, we have some flexibility in the specific form of the objective function, but we require that the objective function $\mathcal{O}$ be a function of $\theta$ and $\langle H \rangle_{U(\theta)}$, expressed as $\mathcal{O} \big( \theta~, \{\langle H \rangle_{U(\theta)}\} \big)$. 

In fact, the selection of the objective function is a critical factor in achieving the desired convergence in VQA. Barren plateaus, where gradients vanish, are closely tied to the choice of the cost function. Therefore, a poor choice can result in a failure to locate a minimum of the objective function through gradient descent. 

\begin{enumerate}
    \item \textbf{Pauli strings}

    The most natural approach to representing our Hamiltonian involves expressing it as a linear combination of basic tensor products of Pauli matrices: $\sigma_i$, with $\sigma_i$ taking values from the set $\{ I, \sigma_x, \sigma_y, \sigma_z \}$. These tensor products are commonly referred to as Pauli strings, defined as $\hat{P} = \otimes_{j = 1}^{n} \sigma_i^{(j)}$, where the index $j$ corresponds to different qubits. This representation is well-suited because the Pauli matrices $\sigma_i$ collectively form a basis for all Hermitian 2 by 2 matrices and serve as the generators of the special unitary group SU(2). The Hamiltonian, then, can be decomposed as $H = \sum_{k=1}^M c_k \hat{P}_k~.$ In this decomposition, the complex coefficient $c_k$ pertains to the k-th Pauli string, and the total number of Pauli strings, denoted as $M$, is contingent upon the specific form of the Hamiltonian under consideration. 
    


    \item \textbf{Fidelity}

    Rather than optimizing with respect to the expectation value of an operator, certain VQAs necessitate a subroutine for optimizing the state obtained from the PQC $U(\theta)$, denoted as $\ket{\Psi}_{U(\theta)}$, with respect to a target state, denoted as $\ket{\Psi}$. We then define fidelity as the following: $F\big(\Psi, \Psi_{U(\theta)}\big) \equiv |\braket{\Psi|\Psi_{U(\theta)}}|^2 = \braket{\operatorname{Proj}_{\Psi}}_{U(\theta)}~,$ which is the expectation value of the projection operator on the target state $\Psi$: $\operatorname{Proj_{\Psi}} = \ketbra{\Psi}{\Psi}~.$ Conventionally, VQA involves minimizing the objective function. In this context, it is natural to use infidelity, represented as $1 - F\big(\Psi, \Psi_{U(\theta)}\big)$, or alternatively, the negative fidelity, denoted as $-\braket{\operatorname{Proj}_{\Psi}}_{U(\theta)}$. 
    
    The expression of objective functions based on fidelities is commonly utilized in state preparation algorithms in quantum optics \cite{kottmann2021quantum,krenn2020computer} and quantum machine learning \cite{benedetti2019generative,cheng2018information,Huang_2021,perez2020data}.
    
\end{enumerate}

Additional objective functions can also be developed, and any cost function expressed operationally can be considered a viable option \cite{bharti2021noisy}. Two examples include the Conditional-Value-at-Risk (CVaR) \cite{barkoutsos2020improving} and the Gibbs objective function \cite{li2020quantum}. These two forms can be simplified to the mean energy, denoted as $\braket{H}~.$ Importantly, their performance is at least as effective as utilizing the mean energy \cite{li2020quantum,barkoutsos2020improving}. The authors of \cite{Lubasch_2020} extend VQE to be applicable to a large class of nonlinear problems, including nonlinear differential equations, and show, for example, that quantum circuits can be exponentially more efficient (in terms of expressivity) than classical matrix product states.

\subsubsection{Parameterized quantum circuits (PQC)} \label{PQC}

Next, we delve into the second critical component of a VQA, which is the quantum circuit responsible for generating the state that optimally satisfies the defined objective. As described above, it prepares the state by an act of unitary that depends on classical parameters: $\theta~,$ hence, it is called the parameterized quantum circuit: PQC. 

The state prepared is defined as the following: $\ket{\Psi(\theta)} = U(\theta) \ket{\Psi_0}~,$ where $\ket{\Psi_0}$ is some initial state, for example, $\ket{0}^{\otimes n}~,$ where $n$ is the number of qubits. As remarked earlier, $n$ should be quite small, due to the limited ability of the NISQ devices, we can only accommodate very shallow circuits. In some VQAs, it is convenient to use other initialization states. We can use a unitary operator (for example, a rotation operator) $P(\phi)$, and the initial state: $\ket{\Psi_0} = P(\phi) \ket{0}^{\otimes n}~.$ 

Similar to the variational principle in quantum wave mechanics, which aims to obtain the wave function with the lowest possible energy, a deep understanding of the physical system can significantly assist us in creating a better initial guess. For instance, if we seek an approximation of the ground state of a potential that exhibits even parity, it's advisable to start with an initial guess that also possesses even parity. It's important to recognize that the choice of the initial state can significantly impact the performance of a VQA in terms of convergence speed and how closely the final state approximates the optimal solution for the problem at hand. This influence is particularly evident in a simple example. Hence, in most cases, your initial step should involve examining the Hamiltonian or the observable of interest to identify any inherent symmetries.

In the NISQ era, we face significant limitations in terms of the number of qubits we can effectively use. Therefore, it's essential to consider the quantum hardware on which our VQA runs. Constructing problem-inspired initial states can be computationally expensive when using native gates. Consequently, most ansätze developed so far fall into one of two categories: problem inspired ansätze and hardware efficient ansätze \cite{bharti2021noisy}. The categorization depends on the structure and intended application of the ansätze. The PQC shown in FIG.~\ref{VQAVague} and FIG.~\ref{VQA4Parts} are examples of the second kind, hardware efficient ansätz. Note that these two circuits are also the same circuit for quantum neural network. We will now discuss both of them below. 

\begin{enumerate}
    \item \textbf{Problem-inspired ansätze}

    In the problem-inspired ansätze approach, we utilize the evolution by a unitary operator, denoted as $G(t)$, with a Hermitian generator $\hat{g}~:$ $G(t) = e^{-i\hat{g}t}.$ These operators are derived from the properties or symmetries of the system under consideration and are used in constructing the PQC. Implementing such unitary transformations directly can be challenging. Therefore, one often employs a technique called the \textit{Suzuki-Trotter (ST) expansion or decomposition} \cite{1976CMaPh..51..183S}. This method provides a general approach to approximate complex and challenging-to-implement unitary transformations. In practice, the more we know about the Hamiltonian to be \textit{Trotterized}, the less number of gates is needed to implement this method \cite{kivlichan2018quantum}. Two mostly used ansätze of this kind is the \textit{unitary coupled cluster} (UCC) for quantum chemistry and the \textit{quantum approximate optimization algorithm} (QAOA). 

    \begin{enumerate}
        \item \textit{Unitary Coupled Cluster (UCC)} 

        Historically, the Unitary Coupled Cluster (UCC) ansätze were among the earliest proposed and implemented approaches, as discussed in reference \cite{Taube2006New}. The core concept behind UCC is to introduce quantum correlations into the Hartree-Fock approximation. Interestingly, in 2014, it was demonstrated that representing UCC on a classical computer is computationally inefficient \cite{Yung2014From}. However, during the same year, researchers took advantage of quantum resources and successfully realized the UCC ansätze as a PQC using a photonic processor \cite{peruzzo2014variational,lee2018generalized,grimsley2019adaptive,bharti2021noisy}.

        \item \textbf{\textit{Quantum Approximate Optimization Algorithm (QAOA)}}

        The Quantum Approximate Optimization Algorithm (QAOA) shares strong connections with quantum annealing and adiabatic quantum computing. Adiabatic quantum computing, in brief, leverages the adiabatic theorem in quantum mechanics, which asserts that a quantum state undergoing slow variations in a time-dependent Hamiltonian will evolve only by a phase. Consequently, one can devise a Hamiltonian that evolves over time, transitioning from an easily prepared system to a system capable of encoding problems to be solved. The energy eigenstate naturally evolves from the initially prepared state to one that solves the target problem. QAOA can be viewed as an approach that resembles Trotterized adiabatic quantum computing \cite{bharti2021noisy,farhi2014quantum}. 

        For combinatorial optimization problems featuring strict constraints, you can introduce penalty terms in the cost function. However, this approach may not be the most efficient in practice since it can still lead to solutions that violate certain hard constraints. A variant of QAOA aimed at addressing these constraints is also discussed in previous works \cite{farhi2014quantum, bharti2021noisy}.

    \end{enumerate}
    
    \item \textbf{Hardware efficient ansätze}

    While computational studies have demonstrated that problem inspired ansätze can yield rapid convergence to a satisfactory solution state, their experimental realization can present challenges. Quantum computing devices are subject to various experimental limitations, including factors like coherence time, limited gate fidelity, and constrained gate set. To give an example on how bad the situation is, \cite{moll2018quantum} illustrates that current quantum hardware is not well-suited for implementing deep and highly connected circuits, which are essential for approaches like the UCC and similar methods designed for applications beyond basic demonstrations, such as simulating more complex molecules like Hydrogen molecule. Human computation capabilities have allowed us to understand the properties of the hydrogen atom and, using manual calculations and the Born-Oppenheimer approximation, we can even compute some characteristics of hydrogen molecules. So, with problem-inspired ansätze on current devices, computers haven't significantly expanded our understanding beyond what we already knew through traditional human-driven calculations.

    Alternatively, a class of hardware-efficient ansätze has been proposed to accommodate device constraints \cite{kandala2017hardware}. The idea is to make good use of a limited number of quantum gates and a particular qubit connection topology. Typically, the gate set used in quantum circuits includes a two-qubit entangling gate and up to three single-qubit gates. The circuit is then built by assembling blocks of single-qubit gates and entangling gates, which are applied in parallel to multiple or all qubits. Each of these blocks is often referred to as a ``layer'', and the ansatz circuit typically comprises multiple such layers.

    The complete VQA program can be broadly conceptualized as a shallow quantum circuit. In this circuit, we iteratively apply rotation gates denoted as $R_{x,y,z}(\theta_{\ell}) = e^{i \theta_{\ell} X_{\ell}}$, with parameters determined by classical values $\theta_k$, followed by the utilization of CNOT gates represented as $W_k$. A basic illustration of this concept is presented in FIG.~\ref{VQAVague}.

    \begin{figure}[!ht]
    \centering
    \includegraphics[width=0.625\textwidth]{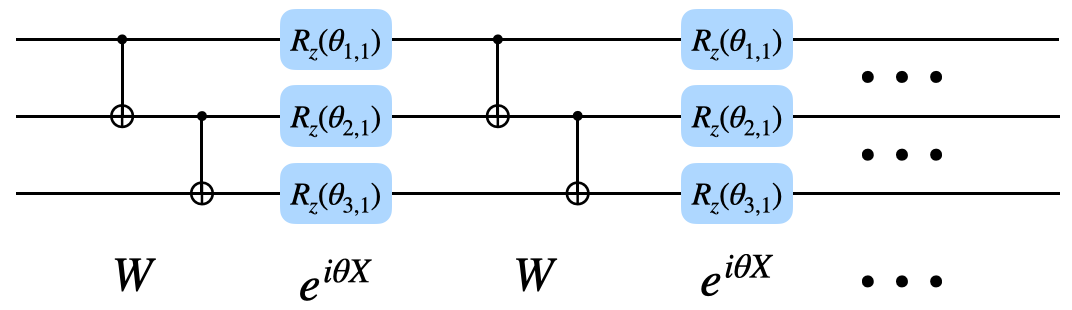}
    \caption{A basic illustration of VQA circuits.}
    \label{VQAVague}
    \centering
    \end{figure}

    Putting all gates together into one unitary operator, we will have 

    \begin{equation} \label{VQAVagueE}
        U = \prod_{\ell=1}^L W_{\ell} U_{\ell}(\theta_{\ell}) = \prod_{\ell=1}^L W_{\ell} \exp \big(i \theta_{\ell} X_{\ell}\big)~.
    \end{equation}
    
    The $\theta_{\ell}$ represent the variational parameters, while $U_{\ell}$ are unitary operators with $X_{\ell}$ denoting the Hermitian operators. Usually, $X_{\ell}$ takes the form of a Pauli operator, and $U_{\ell}$ consists of rotation operators that act on individual qubits. On the other hand, $W_{\ell}$ represents the multi-qubit gates responsible for entangling qubits in the circuit. For example, $W_{\ell}$ might be a CNOT gate, CZ gate, or similar multi-qubit gates commonly used in superconducting qubits \cite{krantz2019quantum}. In line with this methodology, the ``Alternating Layered Ansätze" is a specific instance of hardware-efficient ansätze. It comprises layers of single-qubit rotations and blocks of entangling gates that entangle only a local set of qubits. These entangling gate blocks are shifted every alternating layer in the circuit. The selection of these gates, their interconnections, and their order in the circuit significantly impact the portion of the Hilbert space that the ansatz explores and the rate at which it converges for a given problem \cite{bharti2021noisy}.
    
\end{enumerate}

Instead of strictly choosing between problem-inspired and hardware-efficient ansätze modalities, some PQC designers have adopted an intermediate approach. For instance, they use an exchange-type gate, which can be natively implemented in transmons, to construct a PQC that respects the symmetry of the variational problem. This approach results in ansatz circuits with particularly low parameter counts, making them suitable for quantum chemistry problems like Hydrogen and LiH molecules.

\subsubsection{Measurements} \label{Measurements}

Now, we move into the third stage of VQA, which is the measurement stage. The output of quantum circuits is a quantum state, and in order to optimize it on a classical computer, we must extract classical information from it. This involves estimating the expectation value of the objective function $\braket{\hat{\mathcal{O}}}_{U_{\theta}}$. The most direct approach is to apply a unitary operator to the quantum state, transforming it into the diagonal basis of the observable $\hat{O}$, and then calculate the probability of measuring the state with a corresponding eigenvalue of $\hat{O}$. For experimental details of this approach, see \cite{krantz2019quantum,HAFFNER_2008}. However, this direct approach is limited by the NISQ devices currently available, as transforming the state into the diagonal basis can be computationally expensive. Below, we introduce alternative methods that is more NISQ-friendly.

\begin{enumerate}
    \item \textit{Measurement of Pauli string.} 
    
    Given that most observables can be efficiently expressed as a sum of Pauli strings, a straightforward application is to focus on measuring these Pauli strings. Transforming a state from the computational basis to the diagonal basis of $\sigma_x$ or $\sigma_y$ is relatively simple and can be accomplished using Hadamard gates. 
    
    The Pauli string: $\hat{P} = \prod_{k \in K} \boldsymbol{\sigma}(k)$ where $k$ refers to $k^{th}$ qubit, expectation value is: $ \braket{\hat{P}} = \left\langle \prod_{k \in K} \sigma_z(k) \right\rangle_{\tilde{U}}~,$ where $\tilde{U}$ refers to the product of rotations according to the specific form of the Pauli string. In practice, it's only feasible to take a finite number of single-shot measurements, denoted as $N_s$, of the quantum state. This enables the estimation of expectation values within a finite margin of error. For more information on error of the measurements, see \cite{Huang2019Near}. 

    \item \textit{Measurement of overlaps.} 
    
    Some VQA would require to measure a unitary's overlap with a quantum state $\ket{\psi}~:~~\bra{\psi} U \ket{\psi}~.$ Since this is not an observable, it has both real and imaginary components. The Hadamard test is a method capable of evaluating such a quantity on a quantum computer using just one additional qubit. This involves applying a control qubit with control over the extra qubit and applying the target unitary operator $U$ to the quantum state.

    However, this method does require ample quantum resources to implement the control unitary. An alternative approach, as proposed by the authors of \cite{Mitarai2019Methodology}, involves decomposing $U$ into a sum of Pauli strings and making measurements with respect to these Pauli strings individually. Another approach is applicable when $U$ can be expressed as a product of unitaries, denoted as $U_k$, which act locally on a few qubits. This involves finding classical means to express $U_k$ as $U_k = V_k^{\dagger} D V_k$, where $D$ is a diagonal matrix and $V_k$ represents unitaries that can be applied to $\ket{\psi}$ to determine the overlap. This last approach will lead us naturally to the last part of this section.

    \item \textit{Classical shadows.} 
    
    This technique leverages both the quantum and classical properties of particles. As multiple copies of quantum states' dynamics converge into classical behaviors over time. When dealing with an unknown quantum state represented as $\rho$, the objective is to predict $M$ expectation values, denoted as $\operatorname{Tr}(\hat{O}_i \rho)$, where $1 \leq i \leq M$ \cite{Huang2020Predeicting}.

    This method consists of two steps. Initially, a random unitary transformation, denoted as $U$, is applied to the state $\rho$, resulting in the transformation $\rho \rightarrow U \rho U^{\dagger}$. In the second step, all qubits are measured in the computational basis. These two steps are repeated multiple times using different random unitaries $U$. It's important to note that these random unitaries are selected to be efficiently computable on a classical computer.

    Through post-processing of the measurement results, it is possible to obtain a ``classical shadow," which serves as a classical representation of the quantum state. Research suggests that a classical shadow of size on the order of $\log{M}$ is sufficient to predict all $M$ expectation values simultaneously. This method draws inspiration from the study of shadow tomography \cite{Aaronson2019Shadow}, which will be explored further in section \ref{Shadow}. As mentioned in the introduction, this is an example of the application of other topics in quantum machine learning (shadow tomography) on the narrow sensed quantum machine learning described before. 

\end{enumerate}

\subsubsection{Optimization} \label{Optimization}

Finally, we arrive at the part where we optimize the classical parameters. This optimization process does not differ significantly from other multi-variable optimization problems, so we can employ classical optimization methods \cite{bharti2021noisy}. The challenge here, as mentioned earlier, is that in the NISQ era, we are constrained by the limited coherence time of quantum computers. This means that implementing deep analytical gradient descent circuits may not be feasible.

Additionally, to achieve high precision in the mean value of an observable, a substantial number of measurements are required. Due to the high sampling rate, the measurement process becomes the bottleneck in the overall runtime. Therefore, an ideal optimizer for PQC should aim to minimize the number of measurements needed to be effective. At last, it's crucial for the optimizer to be robust in noisy data environments and maintain precision even with limited measurements taken. 

\begin{enumerate}
    \item \textit{Stochastic gradient descent}

    The technique of gradient descent for locating local minima is a prevalent method in various machine learning applications, particularly in machine learning. While Bayesian learning is intuitively appealing, gradient descent is much more computationally practical.

    The essence of this section can be summarized as follows: when presented with an objective function or loss function, denoted as $\mathcal{L}$, our goal is to identify its minima. The specific form of the loss function is not the primary focus here, but for illustration purposes, one of the most common choices is the mean squared error (MSE) loss:

    \begin{equation} \label{MSL}
        \mathcal{L}_{\text{MSE}} \equiv \frac{1}{2} \left( z(x_{\delta}~;\boldsymbol{\theta}) - y_{\delta} \right)^2~.
    \end{equation}

    In machine learning, the term $z(x_{\delta}~;\boldsymbol{\theta})$ typically represents the final output, and $y_{\delta}$ corresponds to the supervised learning data. In the context of VQA, consider a simple example where $z(x_{\delta}~;\boldsymbol{\theta}) = \braket{\psi}_{U(\theta)}$ and $y_{\delta} = 0$. Therefore, minimizing the loss function is equivalent to minimizing the objective function. All the parameters are encapsulated in the vector $\boldsymbol{\theta}$.

    The core principle of gradient descent is to rapidly approach the minimum when we are far from it and take smaller steps when we are close to the minimum. To determine how far we are from the minimum, we rely on the gradient of the loss function. If the gradient has a large magnitude, it signifies that we are far from the minimum, whereas a small gradient indicates proximity to the minimum. Consequently, we update the parameter values based on the gradient's magnitude:

    \begin{equation} \label{GDEqn}
        \theta_{i} (t+1) \equiv \theta_{i} (t) - \eta~ \frac{\partial \mathcal{L}(\boldsymbol{\theta})}{\partial \theta_{i}} \bigg|_{\theta_{i} = \theta_{i}(t)} ~. 
    \end{equation}

    This update will necessarily decrease the loss function \cite{roberts2022principles}. So, the iterative application of these updates will gradually bring us closer to a local minimum. However, a significant challenge with this method is the need for a substantial number of measurements. To tackle this problem, stochastic gradient descent (SGD) offers a modification to the standard parameter update rule, where it utilizes $\mathcal{L}_{\mathcal{S}_t}$ which serves as an approximation of the loss function. For instance, you can compute the gradient using a finite number of measurements, as described in the work by \cite{Harrow_2021Low}. In machine learning, $\mathcal{L}_{\mathcal{S}_t}$ can be thought of as the loss function computed based on a subset of the training data. These subsets within the training data are randomly divided into equal-sized partitions \cite{roberts2022principles}. 

    \item \textit{Gradient-free optimization}

    There are additional gradient-free optimization approaches that we can explore. Here are a few of these methods, along with their respective references:

    \begin{enumerate}
        \item \textbf{Evolutionary Algorithms}: This technique involves data sampling to estimate the gradient of expected fitness, which is then used to adjust parameters in the direction of steepest improvement \cite{wierstra2011natural,Anand_2021Natual,Zhao_2020Natural}.
    
        \item \textbf{Reinforcement Learning}: In this method, reinforcement learning is initially applied to optimize parameters, as seen in the case of QAOA parameters. It employs a reinforcement learning framework to discover the most effective ``policy" that maximizes the expected total discounted reward \cite{garciasaez2019quantum,yao2020policy,Wauters_2020Reinforcement}.
    
        \item \textbf{Sequential Minimal Optimization (SMO)}: In the field of machine learning, SMO has been highly effective for optimizing the parameter landscape of high-dimensional support vector machines. SMO divides the optimization task into smaller components for which analytical solutions can be derived \cite{platt1998sequential,Nakanishi_2020,Ostaszewski_2021}.
    \end{enumerate}
\end{enumerate}

Also, in \cite{Benedetti_2021}, the authors present variational quantum algorithms for real and imaginary time evolution that are more efficient than previous proposals, where they use Sequential Minimal Optimization for time evolution.

\subsection{Quantum neural tangent kernel} \label{QNTK}

It is important to highlight that the circuits used in quantum machine learning are identical to those employed in variational quantum algorithms (see Fig.~\ref{VQAVague}). Therefore, in this section, we introduce a theory regarding the quantum neural tangent kernel (QNTK), which investigates the dynamics of VQA circuits from first principles \cite{liu2022representation}. This can also be understood as an extension of the first-principle theory for classical machine learning \cite{roberts2022principles}.

Consider an initially prepared state, denoted as  $\ket{\Psi_{0}}$, and apply the unitary operator of the circuit to it (see Eqn.~(\ref{VQAVagueE})): $\ket{\phi(\boldsymbol{\theta})} = U(\boldsymbol{\theta}) \ket{\Psi_0} = \prod_{\ell=1}^L W_{\ell} \exp \big(i \theta_{\ell} X_{\ell}\big) \ket{\Psi_0}~.$ Here, $\ell$ represents the layer of the quantum machine learning system. The output of the quantum neural network is then defined as: $z_{i;\delta} \equiv z_i(\boldsymbol{\theta},\boldsymbol{x}_{\delta}) = \bra{\Psi_0(\boldsymbol{x}_{\delta})} U^{\dagger} \mathcal{O}_i U \ket{\Psi_0(\boldsymbol{x}_{\delta})}~.$ We use the mean squared loss (see Eqn.(\ref{MSL}). We defined the \textit{error factor} or \textit{residual training error} $\varepsilon$ as the part in the parenthesis. The value of $y_{i;\alpha}$ is given by the supervised learning database. 

However, it can be easily set to 0 if one's task is to solve problems like VQE. The gradient descent equation is shown in Eqn.~(\ref{GDEqn}). We then consider the update of the output: 

\begin{equation} \label{QNTKDef}
     \dbar z_{i;\delta} = \dbar \varepsilon_{i;\delta} \equiv \frac{\partial z_{i;\delta}}{\partial \theta_{\ell}} \left( \mathcal{D} \theta_{\ell} \right) = \frac{d z_{i;\delta}}{d \theta_{\ell}} \left(- \eta \frac{\partial \mathcal{L}_{\mathcal{A}}}{\partial \theta_{\ell}} \right) = - \eta \sum_{j,\alpha,\ell} \varepsilon_{i';\alpha} \frac{\partial \varepsilon_{i;\delta}}{\partial \theta_{\ell}} \frac{\partial \varepsilon_{i';\alpha}}{\partial \theta_{\ell}} = - \eta \sum_{j,\alpha} \varepsilon_{i';\alpha} K^{i,i'}_{\delta,\alpha}~,
\end{equation}

The variational angles $\theta_{\ell}$ are then updated step by step, labeled by $t$, according to the derivative of the loss function with respect to $\theta_{\ell}$. $\eta$ is defined as the \textit{learning rate}, which is assumed to be a small (less than 1), positive number. The learning rate determines the size of the step that we take to update our parameters. One can calculate the update of the loss function $\mathcal{L}$ and prove that it decreases indefinitely every time we update our parameters, which is also a fact in classical machine learning \cite{roberts2022principles}. In Eqn.~(\ref{QNTKDef}) it is easy to show that the update of the output is mathematically in the exact same form as the update of the residual training error $\dbar \varepsilon_{i;\delta}$. One then defines the quantum neural tangent kernel (QNTK): 

\begin{equation}
    K^{i,i'}_{\delta,\alpha} \equiv \sum_{\ell} \frac{\partial \varepsilon_{i;\delta}}{\partial \theta_{\ell}} \frac{\partial \varepsilon_{i';\alpha}}{\partial \theta_{\ell}}~.
\end{equation}

If the QNTK is a constant, then Eqn.~(\ref{QNTKDef}) has a closed form solution. In the continuous limit, the residual error $\varepsilon(t)$ will be a decaying exponential with $- \eta K$ as its exponent. So we eventually will have a guaranteed convergence of the output $z_{i;\alpha}$ to the supervised data $y_{i;\alpha}$ if the QNTK $K^{i,i'}_{\delta,\alpha}$ is a constant.

However, the final assumption is not always true. To wit, with the expression of the output \(z_{i,\alpha},\) one can analytically calculate the QNTK and prove that the QNTK is highly fluctuating in the variational angles $\boldsymbol{\theta}$, rather than being a constant \cite{liu2022representation}. This fluctuation implies that we have obtained representation learning. In the terminology of the kernel method (NTK) of classical machine learning, it means that the feature map itself is changing. This indicates that we are extracting features from all of our training data to update our output, and we cannot obtain a closed-form solution for $\varepsilon(t)$ in this case.

However, we can address this problem if we can determine when the QNTK is approximately a constant. If we can achieve this, we will have a partial success and can implement perturbation theory to achieve convergence.

\begin{figure}[!ht]
\centering
\includegraphics[width=0.9\textwidth]{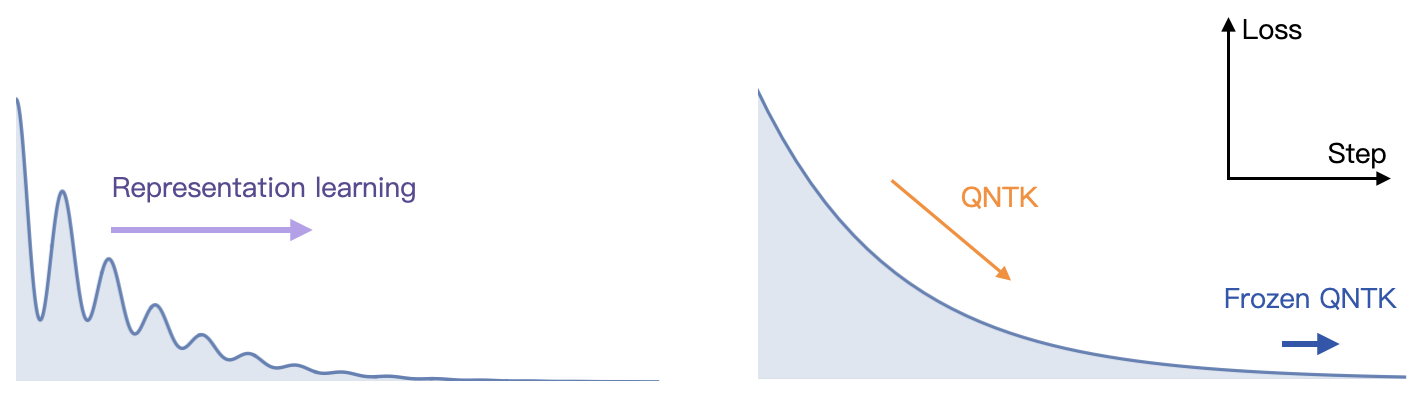}
\caption{Demonstration of the evolution of the loss function in the continuous limit \cite{liu2022representation}. The right diagram depicts the closed-form solution when the QNTK remains constant. In contrast, the left diagram illustrates a scenario where the QNTK is nonlinear but nearly constant. In this case, representation learning is achieved, and the gradient descent equation can be addressed perturbatively.}
\label{QNTKLimit}
\centering
\end{figure}

An instance where the QNTK remains nearly constant is referred to as ``lazy training''. This occurs when the variational angles are nearly unchanged \cite{liu2022representation, chizat2020lazy}: $\theta_{\ell} = \theta_{\ell}^{*} + \delta \varphi~.$ If $\theta_{\ell}$ remains approximately constant, indicating lazy training with minimal movement, the QNTK tends to be nearly constant as well. In such cases, the QNTK expression can be expanded with respect to $\theta_{\ell}$ for predictable dynamics, akin to classical scenarios \cite{roberts2022principles}.

Another scenario involves assuming certain characteristics of the quantum neural network. Firstly, we assume that our variational ansatz $U$ is sufficiently randomly sampled from the unitary group. This enables integration over all samples of $U$ in terms of ensembles, leading to a closed formula for $K$. This assumption is known as the \textit{$k$-design assumption}, commonly applied in various fields such as cryptography and black hole physics. The term ``$k$-design" reflects the prediction accuracy up to the $k^{\text{th}}$ moment compared to the uniform distribution of the unitary group, indicating a completely random choice of $U$.

Upon performing the integral, it is deduced that the average QNTK, denoted as $\overbar{K}$, is proportional to the number of layers $L$:

\begin{equation} \label{AverageQNTK}
    \overbar{K} = \frac{L \operatorname{Tr}(\mathcal{O}_i^2)}{N^2} \propto L~.
\end{equation}

Hence, we can infer that the QNTK increases with the depth of our quantum neural network. Additionally, the standard deviation of QNTK, denoted as $\Delta K$, is proportional to $\sqrt{L}$. This implies that the relative deviation becomes very small as we approach the limit where $L \gg 1$:

\begin{equation}
    \frac{\Delta K}{\overbar{K}} \sim \mathcal{O}(\frac{1}{\sqrt{L}}) \xrightarrow{L \gg 1} \frac{\Delta K}{\overbar{K}} \ll 1~.
\end{equation}

This phenomenon is known as the ``frozen limit," suggesting that if one employs the $k$-design assumption, randomly samples the unitary, and ensures that the quantum neural network is sufficiently deep, a nearly frozen QNTK is obtained. This scenario is commonly referred to as a ``deep quantum neural network."

As $\overbar{K}$ is also proportional to $N^2$ (refer to Eqn.~(\ref{AverageQNTK})), for certain operators such as the Pauli matrices, where the trace scales as $N$, $\overbar{K}$ eventually becomes proportional to $L / N$. This implies that if $L$ is not comparable to the Hilbert space dimension $N$, a small QNTK is obtained, leading to slow convergence. This undermines the practicality of the model, akin to an alternative statement of the barren plateaus phenomena. 

Nonetheless, this model remains useful, as even when not operating at the frozen limit, it provides an approximate means to gauge the speed of gradient descent. For example, to determine which variational ansatz runs the fastest without actually executing them, one can calculate their QNTK and select the one with the largest value. It also enables the estimation of operation time and the number of iterations required for the gradient descent program using Eqn.~(\ref{AverageQNTK}), facilitating hardware design improvements. Moreover, some late-time behaviors of QNTK also been recently captured in \cite{zhang2023dynamical}.

\subsection{Quantum landscapes and barren plateaus} \label{BarrenPlateaus}

Although VQA have been implemented in recent years to address problems in quantum chemistry and similar domains, it remains unclear how well this algorithm will perform in general or under what circumstances VQA might fail. In this section, we introduce barren plateaus as a type of no-go theorem that highlights conditions under which VQA is expected to underperform.

The process of training parameters in the Quantum Machine Learning (QML) model often involves minimizing a loss function and traversing a non-convex landscape in search of its global minimum, as extensively discussed in \cite{cerezo2022challenges}. Quantum landscape theory, as outlined in \cite{Arrasmith_2022}, seeks to comprehend the properties of QML landscapes and explores strategies for engineering them. Notably, local minima and barren plateaus are focal points of investigation within quantum landscape theory

Akin to classical ML \cite{cerezo2022challenges}, the quantum loss landscape may contain numerous local minima. In a manner similar to classical cases, the overall non-convex optimization can become NP-hard \cite{Bittel2021Training}. Various approaches have been proposed to mitigate issues related to local minima. For instance, variable structure Quantum Neural Networks (QNNs), as introduced in \cite{Bilkis_2023,LaRose_2019}, dynamically adjust the model's prior by expanding and contracting during optimization, transforming certain local minima into saddle points. Additionally, there is evidence of the overparametrization phenomenon in QML \cite{kiani2020learning,Larocca_2023}. In this scenario, a computational phase transition occurs as the optimization progresses, leading to the disappearance of spurious local minima when the number of parameters surpasses a critical value. 

Not only are local minima a concern in QML, but there is also the intriguing phenomenon of a \textit{barren plateau} in quantum landscapes \cite{cerezo2022challenges, Cerezo_2021, Arrasmith_2021, Holmes_2022}. A barren plateau is characterized by the loss landscape becoming, on average, exponentially flat concerning the problem size. In this scenario, the valley housing the global minimum exponentially diminishes with the problem size, forming what is known as a \textit{narrow gorge} \cite{Arrasmith_2022}. Consequently, exponential resources, such as an increased number of shots, are needed to traverse through the landscape. This impact on resource requirements complicates the QML algorithm's complexity and can potentially undermine quantum speedup, as quantum algorithms typically aim to avoid the exponential complexity associated with classical algorithms.

The barren plateau phenomenon was first studied in deep hardware-efficient QNNs \cite{McClean_2018}, where they arise due to the high expressivity of the model \cite{Arrasmith_2021, cerezo2022challenges}. By making no assumptions about the underlying data, deep hardware-efficient architectures aim to solve a problem by being able to prepare a wide range of unitary evolutions. In other words, the prior over the hypothesis space is relatively uninformed. Barren plateaus in this unsharp prior are caused by ignorance or the lack of sufficient inductive bias, and therefore a means to avoid them is to input knowledge into the construction of the QNN - making the design of QNNs with good inductive biases for the problem at hand a key solution. Various strategies have been developed to address these barren plateaus, such as clever initialization \cite{verdon2019learning}, pretraining, and parameter correlation \cite{Holmes_2022}. These are also examples of adding a sharper prior to one’s search over the over-expressive parameterizations of hardware-efficient QNNs. The authors of \cite{Cervero_Mart_n_2023} studied the barren plateau problem for parameterized quantum circuits of the quantum tensor network architecture and show, for example, that classical optimization of these circuits can be exponentially more efficient than using a quantum computer.

Other mechanisms have been linked to barren plateaus \cite{cerezo2022challenges}. Simply defining a loss function based on a global observable (i.e., observables measuring all qubits) leads to barren plateaus even for shallow circuits with sharp priors \cite{McClean_2018}, while local observables (those comparing quantum states at the single-qubit level) avoid this issue \cite{McClean_2018,Uvarov_2021}. The latter is not due to bad inductive biases but rather to the fact that comparing objects in exponentially large Hilbert spaces requires an exponential precision, as their overlap is usually exponentially small.

While entanglement is one of the most important quantum resources for information processing tasks in quantum computers, it can also be detrimental for QML models \cite{cerezo2022challenges}. QNNs (or embedding schemes) that generate too much entanglement also lead to barren plateaus \cite{Sharma_2022,marrero2021entanglement,Patti_2021}. Here, the issue arises when one entangles the visible qubits of the QNN (those that one measures at the QNN’s output) with a large number of qubits in the hidden layers. Due to entanglement, the information of the state is stored in non-local correlations across all qubits, and hence the reduced state of the visible qubits concentrates around the maximally mixed state. This type of barren plateau can be solved by taming the entanglement generated across the QNN.

\subsubsection{Influence of background noise}

The existence of hardware noise during quantum computations is a distinctive feature of NISQ computing \cite{cerezo2022challenges}. Despite this reality, many QML studies overlook noise in analytical calculations and numerical simulations, while still asserting near-term compatibility of the methods. It is imperative to consider the impact of hardware noise in QML analyses, especially for those aiming to achieve quantum advantage with currently available hardware \cite{cerezo2022challenges}.

Noise introduces errors in the information as it propagates through a quantum circuit, particularly affecting deeper circuits with longer run-times \cite{cerezo2022challenges}. This impact extends to all aspects of quantum models, including the dataset preparation scheme and circuits used for computing quantum kernels. In the case of Quantum Neural Networks (QNNs), noise can impede their trainability \cite{Wang_2021,wang2021error}, leading to noise-induced barren plateaus where relevant features are exponentially suppressed with increasing circuit depth. The consequence is a deformation of the model's inductive bias and an effective reduction in the dimension of the quantum feature space. Despite the significant influence of quantum noise, its effects remain largely unexplored, especially regarding its impact on the classical simulability of QML models \cite{Deshpande_2022,Hakkaku_2022,cerezo2022challenges}.

Addressing issues induced by noise may necessitate \cite{cerezo2022challenges}: (1) reducing hardware error rates, (2) implementing partial quantum error correction \cite{Bultrini_2023}, or (3) utilizing QNNs that are relatively shallow (i.e., with depth growing sub-linearly in the problem size) \cite{Wang_2021}, such as Quantum Convolutional Neural Networks (QCNNs). While error mitigation techniques \cite{Temme_2017,Czarnik_2021,endo2021hybrid} can enhance the performance of QML models in the presence of noise, they may not completely resolve noise-induced trainability issues \cite{wang2021error}. Another approach involves engineering QML models with noise-resilient properties \cite{LaRose_2020,Cincio_2021,Sharma_2020}, such as ensuring that the position of minima remains unchanged despite noise.

However, another study suggests that noise in VQAs might be beneficial for their performance. Drawing inspiration from classical machine learning, where noise is intentionally introduced to enhance the performance of the gradient descent algorithm, this phenomenon is akin to stochastic gradient descent. It has been observed that gradient descent tends to perform better under the influence of noise, as the noise helps in avoiding saddle points that could hinder the algorithm's progress \cite{Jain_2017, jin2019nonconvex}. In the absence of noise, the algorithm may converge directly to the saddle point. However, when noise is introduced, perturbations around the points prevent convergence towards these unstable fixed points.

In the study \cite{liu2022noise}, this discussion was extended to quantum noise. Through experiments, the researchers found that the noisy case outperformed the noiseless counterparts, providing a toy model illustration to support their argument. This suggests that, while implementing quantum error correction is crucial for making quantum devices fault-tolerant, in certain QML applications, eliminating all noise may not be necessary. As long as the noise remains below a certain threshold, it may prove not detrimental but rather helpful for QML programs.

\subsubsection{Laziness}

Theoretically one can understand barren-plateau phenomenon as the following. A typical gradient descent algorithm will look like 

\begin{equation} \label{laziness}
    \theta_{\ell}(t+1) - \theta_{\ell}(t) = - \eta \frac{\partial \mathcal{L}}{\partial \theta_{\ell}} \equiv \delta \theta_{\ell}~.
\end{equation}

The observation \cite{McClean_2018} is that if our variational ansatz is highly random, according to the $k$-design integral formula \cite{Roberts_2017, Liu_2020, Liu_2018, Cotler_2017}, the derivative of the loss function is generally suppressed by the dimension of the Hilbert space $N$. This might lead to a situation where the variation of the loss function during gradient descent is very small, denoted as $\delta \mathcal{L} \equiv \mathcal{L}(t+1) - \mathcal{L}(t) \ll 1$ for the step $t$. An alternative theoretical term used to describe the quantum barren plateau, based on the large suppression in Eqn.~(\ref{laziness}), is laziness \cite{liu2022laziness}. In the quantum context, the suppression comes from the dimension of the Hilbert space, while in the classical case, the suppression is associated with the width of classical neural networks.

To be more precise, laziness refers to small $\delta \theta_{\mu}$, and the barren plateau refers to small $\delta \mathcal{L}$. The authors of \cite{liu2022laziness} demonstrate that laziness may not necessarily imply the quantum barren plateau. This insight is derived from the perspectives of overparametrization theory and representation learning theory through the QNTK \cite{liu2022representation}. In the context of quantum neural networks, \textit{overparametrization} refers to the condition $\overbar{K} \approx \mathcal{O}(1)$ in Eqn.(\ref{AverageQNTK}). 

The paper \cite{liu2022laziness} emphasizes that in variational circuits with a sufficiently large number of trainable angles, gradient descent dynamics can still be efficiently executed, even in the presence of the exponential suppression of variational angle updates (laziness). On the other hand, the only limitation is the precision, where noises might affect the performance significantly. The authors also highlight that laziness is not exclusive to quantum machine learning but is observed in overparametrized classical neural networks with large widths as well. The efficiency of large-width neural networks is substantiated by the neural tangent kernel theory, and similar principles apply to their quantum counterparts.

\section{Fault tolerant quantum computation (FTQC) algorithms} \label{FTQC}

In the Fault-Tolerant Quantum Computation (FTQC) era, we envision a future where we have fully implemented quantum error correction. In this era, one of the most prominent and impactful applications is Shor's algorithm \cite{Shor1994Algorithm}. This algorithm has the remarkable ability to efficiently factorize large numbers, a capability that could potentially break the majority of the public-key cryptography systems currently used on the internet. Additionally, quantum computers, as originally envisioned by Richard Feynman \cite{feynman1996lectures}, have the potential to significantly accelerate the simulation of quantum physics and chemistry.

Despite these two remarkable applications, quantum computers have historically faced a problem of having limited use cases. While these two applications are groundbreaking, there is no concrete evidence that quantum computers will revolutionize our lives in the same way that classical computers have done \cite{aaronson2015read}.

Recently, a series of newly proposed quantum algorithms have emerged, offering exciting possibilities for quantum computation. Among them, the Harrow-Hassidim-Lloyd (HHL) algorithm, introduced by its authors in 2008 \cite{harrow2009quantum}, has gained significant attention. These novel algorithms not only hold the potential for significant speedups compared to classical counterparts but also show promise in addressing practical problems. These applications span various domains, including machine learning, clustering, classification, and the analysis of extensive datasets \cite{aaronson2015read}. Although, significant challenges are related to many algorithms designed for quantum machine learning, including dequantization and fast interfaces between classical and quantum processors. 

In this section, we will concentrate on a few selected quantum algorithms and delve into their applications. 


\subsection{Quantum phase estimation and quantum principle component analysis} \label{QPE&qPCA}

In the realm of quantum computing, quantum phase estimation is a pivotal quantum algorithm primarily designed to estimate the phase associated with an eigenvalue of a provided unitary operator \cite{nielsen_chuang_2010}. The necessity to exponentiate the density matrix within quantum principle component analysis (PCA) aligns with the requirement for phase estimation.

\begin{figure}[!ht]
\centering
\includegraphics[width=0.5\textwidth]{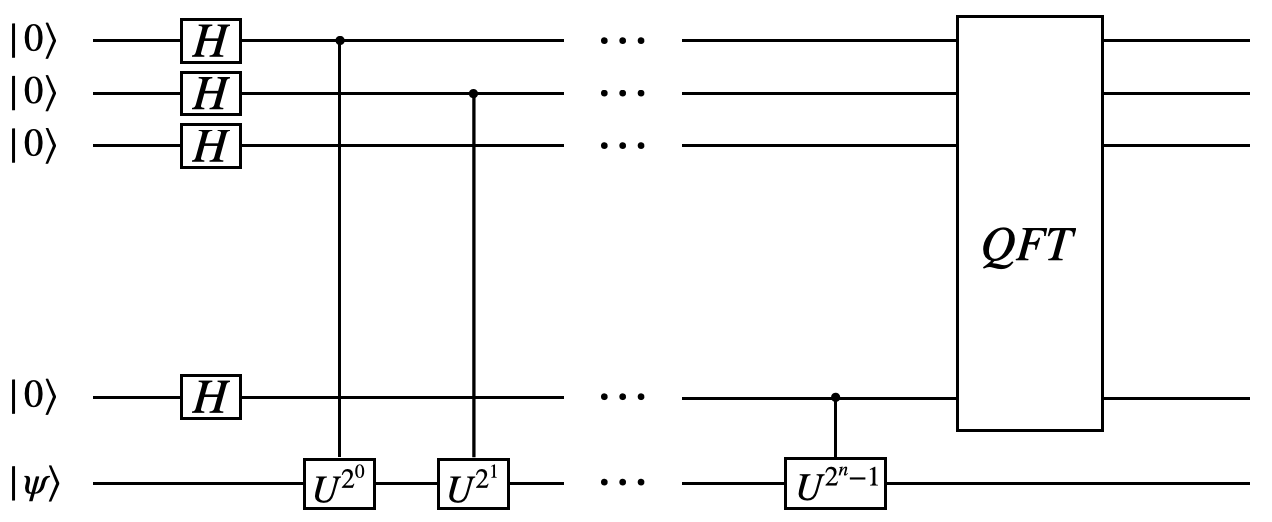}
\caption{Circuits for quantum phase estimation.}
\label{QPEf}
\centering
\end{figure}

In a formal context, suppose a unitary operator $U$ and its corresponding eigenvector $\ket{\psi}~.$ The relationship is represented as: $U \ket{\psi} = e^{i 2 \pi \theta} \ket{\psi}~.$ Here, without loss of generality, we assume $0 \leq \theta < 1~.$ and the objective of Quantum Phase Estimation (QPE) is to accurately determine the value of $\theta$ while minimizing the number of operations required. The concept involves generating a binary approximation of $\theta$ through the application of quantum Fourier transformation. The preparatory steps are illustrated in FIG.~\ref{QPEf}.

Beginning with $\ket{\psi}$ and $n$ qubits initialized in the $\ket{0}$ state: $\ket{0}^{\otimes n} \otimes \ket{\psi}~.$ First one applies Hadamard gates to all $\ket{0}$ qubits: $H^{\otimes n} \ket{0}^{\otimes n} = \frac{1}{2^{n/2}} \left( \ket{0} + \ket{1} \right)^{\otimes n} = \frac{1}{2^{n/2}} \sum_{k=0}^{2^{n-1}} \ket{k}~.$ One then decomposes $k~:$ $k = \sum_{j=0}^{n-1} k_j 2^{j}~.$ The next step involves applying the unitary operator $U$ to $\ket{\psi}$ while maintaining control over the other qubits. Practically, one applies $U^{2^{j}}$ on $\ket{\psi}~,$ only if $k_j=1~,$ for $0 \leq j \leq n-1~.$ Finally, this sequence of operations results in the state:

\begin{equation}
    \frac{1}{2^{n/2}} \sum_{k=0}^{2^{n-1}} \ket{k} \ket{\psi} \rightarrow \frac{1}{2^{n/2}} \sum_{k=0}^{2^{n-1}} \ket{k} U^{k} \ket{\psi} = \ket{k} U^{k_0 2^{0}} U^{k_1 2^{1}} \ldots U^{k_{n-1} 2^{n-1}} \ket{\psi} = \frac{1}{2^{n/2}} \sum_{k=0}^{2^{n-1}} e^{i 2 \pi k \theta} \ket{k} \ket{\psi}~.
\end{equation}

The eigenstate $\ket{\psi}$ is not needed at this stage. We proceed with the quantum Fourier transformation on the qubits $\ket{k}~:$

\begin{equation}
    \begin{split}
        \frac{1}{2^{n/2}} \sum_{k=0}^{2^{n-1}} e^{i 2 \pi k \theta} \ket{k} = \frac{1}{2^n} \sum_{x=0}^{2^n-1} \sum_{k=0}^{2^{n-1}} \exp \left\{\frac{- i 2 \pi k}{2^n} \left( x - 2^n \theta \right) \right\} \ket{x}~.
    \end{split}
\end{equation}

Now, one estimates $\theta$ by rounding $2^n \theta$ to the nearest integer. If $2^n \theta$ can be expressed as an integer, then one will obtain a precise result after performing measurements. On the other hand, If $\theta$ cannot be accurately expressed up to $n^{th}$ order binary, QPE will have an error. So the larger $n$ is, the smaller the error is. Because we have more space to approximate $\theta~.$ The good news is the error of QPE estimation is always bounded \cite{nielsen_chuang_2010}. By approximating $\theta$, it's possible to find the eigenvalue of the unitary operator. Consequently, since unitary operators can always be expressed as an exponential of a Hermitian operator, one can also determine the eigenvalue of the Hermitian matrix. This is the principle behind how quantum PCA calculates both the eigenvectors and eigenvalues of the density matrix.

One of the most fascinating applications in quantum machine learning is Quantum Principal Component Analysis. This method, akin to the popular Principal Component Analysis (PCA) used by companies like Netflix, enables the analysis of users' preferences to offer tailored film recommendations. In a mathematical sense, the data provided by users is represented as vectors: $v_j~,$ in a $d$ dimensional vector space. PCA's objective is to identify the ideal axes for grouping this data into clusters. Another application could involve generating these vectors based on the fluctuations in stock prices for all stocks in the market, analyzing changes from discrete times $t_i$ to $t_{i+1}~$ \cite{biamonte2017quantum}. 

\begin{figure}[!ht]
\centering
\includegraphics[width=0.3\textwidth]{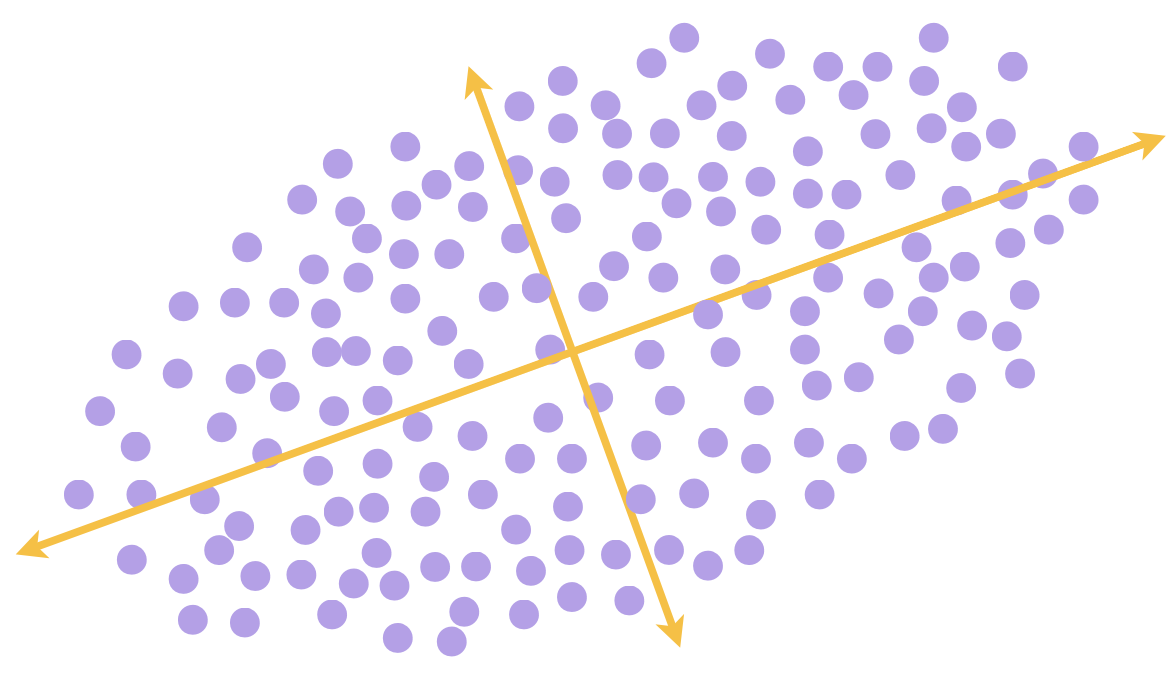}
\caption{A simple illustration of principle component analysis (PCA).}
\label{Blob}
\centering
\end{figure}

Classically, this program is done as the following \cite{biamonte2017quantum}. First, one calculate the covariance matrix: $ \boldsymbol{C} = \sum_j \boldsymbol{v}_j \boldsymbol{v}_j^T~,$ where $T$ represents the transverse matrix. In principle, one can understand the covariance matrix as a manifestation of the correlation between different component of the data. Now, one can diagonalize the covariance matrix and find its eigenvectors $c_k$ and corresponding eigenvalue $e_k$~: $\boldsymbol{C} = \sum_k e_k \boldsymbol{c}_k \boldsymbol{c}_k^{\dagger}~.$

If only a few of the eigenvalues $c_k$ are large, and the remainder are small or zero, then the eigenvectors corresponding to those eigenvalues are called the principal components of $\boldsymbol{C}~.$ In more mathematical terms, if the diagonalized covariance matrix is sparse, then the non-zero eigenvalues are called the principle components. Each principal component represents an underlying common trend or form of correlation in the data, and decomposing a data vectors $\boldsymbol{v}$ in terms of principal components,

If only a few of the eigenvalues $c_k$ are significantly large, and the rest are considerably small or zero, then the eigenvectors linked to these eigenvalues are termed the principal components of $\boldsymbol{C}$ In more technical terms, if the diagonalized covariance matrix is sparse, the non-zero eigenvalues represent the principal components. Each principal component characterizes an underlying common trend or form of correlation within the data. Decomposing a data vector $\boldsymbol{v}$ into these principal components through the equation: $\boldsymbol{v} = \sum_{k} \tilde{v}_k  \boldsymbol{c}_k~,$ enables compression of data representation and the anticipation of future behavior. Classical algorithms for PCA possess computational complexity and query complexity scaling as $\mathcal{O}(d^2)~.$

In the quantum version of this program, the initial step involves mapping these vectors $\boldsymbol{v}_j$ into quantum states: $\ket{v_j}~.$ This mapping is facilitated by quantum random access memory (QRAM), a concept that will be further detailed in section \ref{QRAM}. A $d$ dimensional vector can be encoded in $\log{d}$ qubits. Assuming every vector $\boldsymbol{v}_j$ is equally probable or is randomly selected from a dataset containing $N$ vectors the resulting density matrix is: $\rho = \frac{1}{N} \sum_{j} \ketbra{v_j}{v_j}~.$ This density matrix essentially serves as the quantum equivalent of the covariance matrix $\boldsymbol{C}$ in the classical scenario. The subsequent step for quantum principal component analysis involves implementing two methods: density matrix exponentiation and quantum phase estimation. The former maps the density matrix $\rho$ into a unitary operator: $U_{\text{DME}} = e^{- i \theta \rho}~,$ which enables us to utilize QPE to evaluate the eigenvalues and eigenstates.  


The latter method aims to find the eigenvectors and eigenvalues of the density matrix written in $U_{\text{DME}}$. By repeatedly sampling data and using the two aforementioned techniques, one can take any quantum version of a data vector $\ket{\boldsymbol{v}_j}$ and decompose it into its principal components: $\ket{v_j} \rightarrow \sum_{k} v_k \ket{c_k}~.$ It's important to note that quantum principle component analysis scales as $\mathcal{O}(\log^2{d})$ in both computational complexity and query complexity, which represents a significant speedup compared to the classical version. In the study conducted in \cite{huang2022quantum}, the researchers present experimental evidence showcasing quantum advantage in PCA. However, a crucial prerequisite for this advantage is the availability of quantum data sourced from physical experiments, capable of generating density matrices. This contrasts with classical devices relying on QRAM for data processing. And the quantum advantage is measured in the number of queries to get the density matrix.


\subsection{Classical algorithm framework for dequantizating QML models} \label{Dequantization}

Despite the development of exciting quantum algorithms in QML, there are still open problems whether known QML algorithms provide new exponential speedups over classical algorithms for practically relevant instances of machine learning problems \cite{Ciliberto_2018,Dunjko2020Non}. This uncertainty is rooted in the challenge of identifying the source of this speedup. Some arguments suggest that QML's exponential acceleration is solely attributed to its state preparation assumptions. Notably, the concept of ``dequantization" has emerged, involving the development of classical algorithms that mimic the sampling assumptions of QML programs. Authors in this domain have created classical algorithms, often referred to as the dequantization of QML programs, which closely resemble their quantum counterparts but exhibit only polynomial differences in complexity. One notable example of dequantization involves the quantum recommendation system.

The work by Kerenidis and Prakash on the quantum recommendation system \cite{kerenidis2016quantum} was noteworthy for addressing the caveats associated with the HHL algorithm, as pointed out by Scott Aaronson \cite{aaronson2015read}. They presented a comprehensive quantum algorithm directly comparable to classical algorithms. Initially, Kerenidis and Prakash's quantum algorithm demonstrated exponential speedup compared to the best-known classical algorithms, but its provable exponential nature was uncertain. In 2019, Tang \cite{tang2019quantum} introduced a classical algorithm that dequantized the quantum recommendation systems algorithm with only polynomially slower runtime, providing clarity on the provable nature of the speedup. These works highlighted the insight that the data structure fulfilling state preparation assumptions can also fulfill $\ell^2$-norm sampling assumptions (as defined in section 2.2 of \cite{tang2019quantum}). More precisely, this is a $\log$-dimension time classical randomized linear algebra algorithms in the sample and query input model, which is an analogue of quantum state preparation assumptions coming from, for example, in most cases, QRAM). Consequently, a classical algorithm aiming to ``match" the quantum algorithm can leverage these assumptions. 

In 2020, the authors of \cite{Chia_2020} introduced an algorithmic framework for quantum-inspired classical algorithms focusing on close-to-low-rank matrices. This work generalized the series of results initiated by Tang's quantum-inspired algorithm for recommendation systems \cite{tang2019quantum}. The motivation for these classical algorithms stemmed from quantum linear algebra algorithms and the quantum singular value transformation (SVT) framework \cite{Gily_n_2019}. The proposed classical algorithms for SVT exhibited runtime independent of input dimension under suitable quantum-inspired sampling assumptions. The results provided compelling evidence that, in the corresponding QRAM data structure input model, quantum SVT does not deliver exponential quantum speedups.

In 2021, the authors of \cite{Tang_2021} introduced a new input model termed \textit{SQ access}, representing a form of l2-norm sampling assumption. They developed a model to dequantize two influential QML algorithms, namely quantum principal component analysis \cite{Lloyd_2014} and quantum supervised clustering \cite{lloyd2013quantum}. In essence, they provided classical algorithms that, with classical SQ access assumptions replacing quantum state preparation assumptions, replicated the bounds and runtime of the corresponding quantum algorithms, albeit with a polynomial slowdown. 

On the other hand, in \cite{huang2022quantum} the authors point out that the assumption in \cite{Tang_2021} does not work in the quantum physics environment setup, and it is possible that there are still quantum advantages with exponential separation. In \cite{huang2022quantum}, their quantum advantages are measured in terms of the number of queries to access the quantum state requiring principle component analysis. In \cite{Tang_2021}, It presupposes the capability to retrieve any element of the exponentially large density matrix with exponentially precise accuracy, within a time frame that is polynomial. Achieving the capability already requires exponential numbers of queries to access the density matrix. See further detailed discussions in \cite{cotler2021revisiting}. In summary, the pursuit of identifying classical analogs for quantum machine learning applications presents a fascinating avenue for exploration. Investigating dequantization enables the identification of the constraints inherent in both quantum and classical machine learning, and aids in delineating the distinctions between quantum and classical algorithms.

\subsection{Harrow-Hassidim-Lloyd algorithm} \label{HHL}

In this section, we will delve into the intricacies of the HHL algorithm, as detailed in \cite{harrow2009quantum}. The fundamental setup remains consistent:

\begin{equation} \label{HHLDEqn}
    A \ket{x} = \ket{b}~,
\end{equation}
where $A$ is an $N \times N$ Hermitian matrix, and $\ket{x}$ and $\ket{b}$ are normalized vectors. We assume that $A$ is Hermitian which does not compromise the generality of the algorithm, as the space can always be expanded to make it true. For $\ket{u_i}$ the eigen vectors of $A$, we can decompose $A$ and $\ket{b}$, such that 

\begin{equation} \label{HHLE}
    \ket{x} = A^{-1} \ket{b} = \sum_{i,j=0}^{N-1} \lambda_i^{-1} \ketbra{u_i}{u_i} b_j \ket{u_j} = \sum_{i,j=0}^{N-1} \lambda^{-1}_{i} b_j \delta_{ij} \ket{u_i} = \sum_{i=0}^{N-1} \lambda^{-1}_{i} b_i \ket{u_i}~.
\end{equation}

Now, let's delve into the HHL algorithm circuit, which step by step accomplishes what we've discussed in Eqn.~(\ref{HHLE}). We'll track the evolution of qubit states as follows.

\begin{figure}[!ht]
\centering
\includegraphics[width=0.7\textwidth]{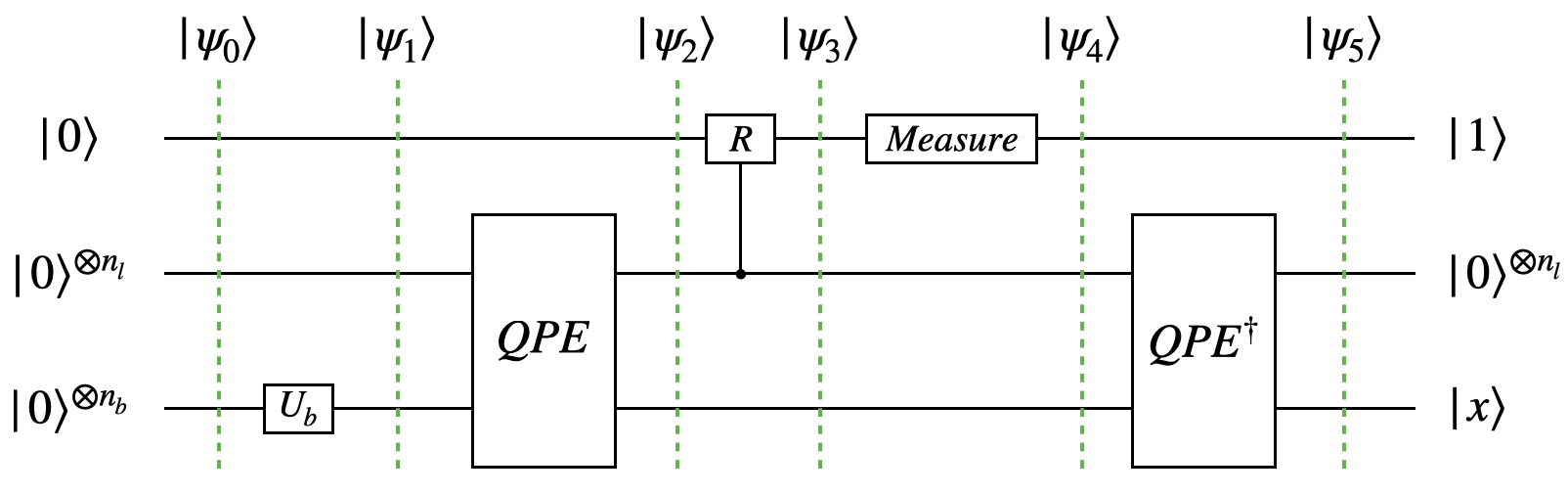}
\caption{Circuits for HHL algorithm \cite{morrell2023stepbystep}.}
\label{HHLFig}
\centering
\end{figure}

With the entire process divided into 5 stages \cite{harrow2009quantum,morrell2023stepbystep}, we begin with the following initialization: $\ket{\psi_0} = \ket{0}^{\otimes n_b} \ket{0}^{\otimes n_l} \ket{0}~.$

The goal is to encode the information of $\ket{b}$ in the first register: $\ket{0}^{\otimes n_b}$ and encode $\lambda_j$ in the second register: $\ket{0}^{\otimes n_l}~.$ These can be done with a unitary operator $U_b$ and quantum phase estimation which determines the values of $\lambda_j$ and $\ket{u_j}~:$ $\ket{\psi_2} = \sum_{j = 0}^{N-1} b_j \ket{u_j} \ket{\lambda_j} \ket{0}~.$

The key step is to rotate the third register qubit with an angle with a control on the clock register and do a post-selection on the third register of obtaining $\ket{1}$:

\begin{equation}
    \ket{\psi_3} = \sum_{j = 0}^{N-1} b_j \ket{u_j} \ket{\lambda_j} \left( \sqrt{1-\frac{C^2}{\lambda_j^2}} \ket{0} + \frac{C}{\lambda_j} \ket{1} \right) \xrightarrow{\text{post-select}} \ket{\psi_4} = \sum_{j = 0}^{N-1} b_j \ket{u_j} \ket{\lambda_j}  \frac{C}{\lambda_j} \ket{1}~,
\end{equation}
where constant $C$ is the normalization factor. If we obtain $\ket{0}~,$ hen we will have to repeat the above procedure until we obtain the desired result. Now we have our desired $\lambda_j^{-1}$, so we can apply the inverse of the quantum phase estimation procedure and return to our initialized second register and we have the following identification:

\begin{equation}
    \ket{\psi_5} = C \sum_{j = 0}^{N-1} b_j \lambda_j^{-1} \ket{u_j} \ket{0}^{\otimes n_l} \ket{1} = C \ket{x} \ket{0}^{\otimes n_l} \ket{1}~.
\end{equation}

The state output of the first register is our desired solution state or vector. The whole post-selection process is for us to be sure that the state in the first register is now proportional to the solution vector, and from there, we can complete our algorithm. For a more detailed discussion of the HHL algorithm, please refer to the original paper \cite{harrow2009quantum}. The HHL algorithm takes $\mathcal{O}((\log{N})^2)$ quantum steps to output $\ket{x}$, compared with $\mathcal{O}(N \log{N} )$ steps required to find the vector $\boldsymbol{x}$ using the best-known method on a classical computer. HHL does achieve a significant speedup, and so does the quantum SVM compared to its classical counterpart.

Before we become overly enthusiastic about this algorithm and its broad applications, it's important to note the following caveats, which can be crucial in practice \cite{aaronson2015read}.

\begin{enumerate}
    \item HHL can handle various types of information. However, when dealing with classical information, it's essential to efficiently load the vector $b = (b_1, \ldots, b_n)$ into the quantum computer's memory. This process involves transforming:

    \begin{equation}
        \boldsymbol{b} \rightarrow \ket{b} = \sum_{i=0}^{N-1} b_i \ket{u_i}~,
    \end{equation}

    must occur rapidly. If the preparation of the state $\ket{b}$ requires exponential steps, the significant speedup gained by HHL in subsequent steps might be negated. Theoretically, this might be achieved through quantum random access memory (QRAM), which will be discussed in section \ref{QRAM}. Nevertheless, it is crucial that $\boldsymbol{b}$ is reasonably uniform. For evident reasons, this challenge is referred to as the input problem \cite{aaronson2015read, biamonte2017quantum}.
    
    \item The quantum computer must also have the capability to apply unitary transformations of the form $e^{-iAt}$ for various values of $t$. If the matrix $A$ is \textit{sparse}—meaning it contains at most $s$ nonzero entries per row, where $s \ll n$—and if there exists a quantum random access memory (QRAM) that efficiently stores, for each $i$, the locations and values of row $i$’s nonzero entries, then it is known that one can apply $e^{-iAt}$ in a time that grows nearly linearly with $s$.

    \item The matrix $A$ not only needs to be invertible but also robustly invertible, or ``well-conditioned." Otherwise, the significant speedup provided by the algorithm may be lost.

    \item Finally, there's the output problem, which is the converse of the input problem \cite{biamonte2017quantum}. The challenge of encoding classical information into quantum states also extends to reading the output quantum state $\ket{x}$ into classical information $\boldsymbol{x}$ when HHL is finished. Learning the value of any specific entry $x_i$ by measurement will, in general, require repeating the algorithm roughly $N$ times, negating the significant speedup.
\end{enumerate}

To summarize, HHL is an algorithm designed to solve a system of linear equations approximately in logarithmic time. It serves as a template for other quantum algorithms, as long as one can address and mitigate the various caveats associated with its practical implementation \cite{aaronson2015read}. If these challenges are properly handled, HHL can find applications in various real-world scenarios.

\subsubsection{Large scale machine learning models in fault tolerant era} \label{Carleman}

There are trials about applications of the HHL algorithm and their implementation, particularly at a large scale, in Quantum Machine Learning (QML) programs. One observation is that, the mathematical description of the gradient descent program involves solving a set of difference equations, which, in the continuous limit, transforms into ordinary differential equations (ODEs).

The paper \cite{Liu2021Efficient} introduces a technique known as \textbf{Carleman linearization}, demonstrating that non-linear ODEs can be linearized into an infinite set of linear ODEs that can be truncated to a chosen level of approximation. This allows the approximate solution of challenging non-linear ODEs by solving the easier linearized versions. The authors establish that if the original ODEs are non-linear and \textit{dissipative}, then the error resulting from the truncation in Carleman linearization can be controlled. This implies that a non-linear, dissipative system can be efficiently linearized.

In the realm of quantum computing, the HHL algorithm is known to efficiently solve linear ODEs, as expressed in Eqn.~(\ref{HHLDEqn}), providing a significant speedup compared to classical algorithms. Consequently, the combination of Carleman linearization and the HHL algorithm holds the promise of yielding a significantly efficient algorithm for solving non-linear ODEs in comparison to classical algorithms. 

Dissipativeness plays a crucial role in various machine learning tasks. Many machine learning processes, including gradient descent, exhibit dissipative behavior. Treating gradient descent as dynamics reveals that it operates as an open first-order system outside the scope of Lagrangian mechanics \cite{liu2023towards}. Biological processes in the brain also exemplify the significance of dissipation, where forgetting becomes essential for making room to remember new information \cite{liu2023towards}.

The findings of \cite{liu2023towards} suggest the potential for an efficient algorithm to solve stochastic gradient descent. Generalizing the techniques for differential equations to \textit{ordinary difference equations}, it is observed that in the learning process, the Hessian eigenvalues characterize the degree of dissipation. Towards the end of training, there are roughly equal numbers of positive and negative eigenvalues, signifying comparable dissipative and non-dissipative modes. However, at the beginning of training, the number of dissipative modes significantly exceeds the number of non-dissipative modes, creating highly asymmetric Hessian eigenvalues. This phase is recognized as highly dissipative, consistent with the intuition that the system learns rapidly in the initial stages. The ODE solver described earlier is applicable during this phase, enabling the solution of machine learning tasks. Towards the end of training, when positive and negative eigenvalues become comparable, the algorithm may not be suitable for solving ODEs in general. An exception is when approaching a local minimum, where the Hessian matrix is positive definite, indicating a dissipative system suitable for the ODE solver. In summary, this quantum algorithm might help machine learning processes significantly, and if the Hessian becomes symmetric towards the end, classical algorithms can be employed for further computation. 

Overcoming the input and output challenges in quantum machine learning is a complex task. While Quantum Random Access Memory (QRAM) is a promising solution for efficient data input, it is still in the developmental stage. Addressing the input challenge in quantum machine learning can be tackled using a classical algorithm known as \textit{``pruning"}. Typically, towards the end of training in many machine learning algorithms, the matrix of training parameters becomes highly sparse, meaning that a significant number of training parameters become zero. This sparsity presents an opportunity to effectively manage the input problem in quantum machine learning. 

Dealing with the output problem involves leveraging the sparsity that often occurs in parameter matrices at the end of training, making it potentially easier to download sparse quantum states to classical computers. Additionally, classical shadow techniques, with their polynomial scaling, provide a method for extracting information from dense quantum states, if we know the sparsity of the matrix. These areas are actively researched, and advancements in quantum computing technologies and algorithms may offer new solutions in the future.

Indeed, the vision of leveraging quantum algorithms, particularly with large-scale implementations of algorithms like the HHL, holds the promise of addressing the increasing costs associated with classical algorithms. The potential gains in efficiency and computational speed offered by quantum computers could revolutionize machine learning and other computational tasks. However, it's important to emphasize that realizing these advancements requires the implementation of full quantum error correction to ensure fault tolerance. Overcoming the challenges associated with error correction remains a critical aspect of bringing quantum computing applications to fruition. Moreover, some further explorations about machine learning foundations should be given along this line, like improving the algorithms from solving gradient descent to solving the matrix multiplication problems in large-scale classical neural networks for back propagation, where quantum linear algebra techniques or parallel computing from analog photonic devices might be helpful for those tasks \cite{becker2023optoacoustic}.

\subsection{Quantum random access memory} \label{QRAM}

Now, let's explore the challenges associated with the ``input problem" and the ``output problem". Classical data needs to be inputted into a quantum computer before undergoing processing \cite{biamonte2017quantum}. Referred to as the `input problem', this step is generally performed with minimal overhead but can pose a significant bottleneck for certain algorithms. Similarly, the `output problem' arises when retrieving data after it has been processed on a quantum device. Much like the input problem, the output problem can lead to a substantial operational slowdown. 

In particular, when considering the application of algorithms like HHL, least squares fitting, quantum support vector machines, and related approaches, substantial amounts of classical data might be loaded into a quantum system. This loading process can demand significant time, as acknowledged in the literature \cite{aaronson2015read,biamonte2017quantum}. While the use of a quantum random access memory (QRAM) \cite{hann2021practicality,Hann1} might, in principle, address this issue, the associated costs might be prohibitive for large-scale data problems \cite{Arunachalam_2015}. 

QRAM can be conceptualized as an architecture designed for the implementation of quantum \textit{oracles} \cite{hann2021practicality}. Consider a computational problem where the input is represented by a classical data vector $\boldsymbol{x}~.$ An oracle, often described as a black box, is a tool that can be interrogated to disclose information about $\boldsymbol{x}~.$ Although the oracle reveals details about $\boldsymbol{x}$ upon request, the specific method it employs to obtain this information is not explicitly specified. 

\begin{figure}[!ht]
\centering
\includegraphics[width=0.9\textwidth]{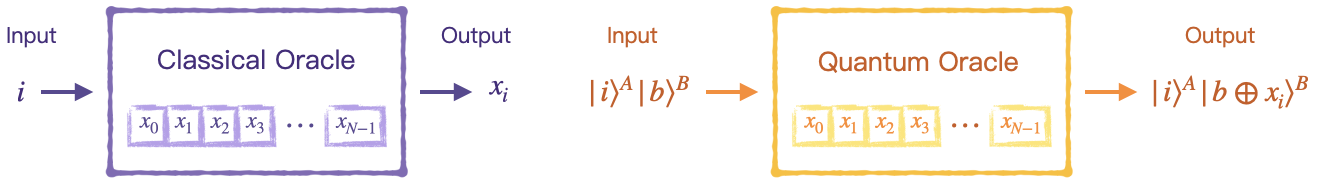}
\caption{Classical and quantum data-lookup oracles \cite{hann2021practicality}.}
\label{Oracle}
\centering
\end{figure}

Oracles can exist in either classical or quantum forms, as illustrated in FIG.~\ref{Oracle}. A basic example of a classical oracle is the data-lookup oracle. This type of oracle is activated by providing it with an index $i$ as input, and it subsequently outputs the corresponding vector element $x_i~.$ A natural extension of this classical data-lookup oracle is the quantum data-lookup oracle. In the quantum scenario, both the inputs and outputs of the oracle query are represented as quantum states. The query itself is implemented by a unitary operation, denoted as $\mathcal{O}^{(\text{DL})}_x$, which performs the mapping \cite{hann2021practicality}:

\begin{equation}
    \mathcal{O}^{\text{DL})}_x \ket{i}^A \ket{b}^B = \ket{i}^A \ket{b \oplus x_i}^B~,
\end{equation}

where the notation $b$ represents any computational basis state and $\oplus$ denotes addition modulo 2. The superscripts $A$ and $B$ refer to two quantum registers, the state of register $A$ indicates which element to look up, and the query encodes this element into the state of register $B~.$ It's important to note that, while the inputs and outputs of the query are quantum, the data being queried is classical. This characteristic makes quantum oracles serve as a bridge between classical data and quantum algorithms \cite{hann2021practicality}. 

However, merely loading classical data into quantum memory is not sufficient; the process needs to be fast \cite{aaronson2015read}. QRAM is designed to address this requirement. In QRAM, the quantum state summarizing the vector uses $\log{d}$ qubits, and the QRAM operation involves $\mathcal{O}(d)$ operations distributed over $\mathcal{O}(\log{d})$ steps, which can be performed in parallel \cite{biamonte2017quantum}.

For QRAM, a quantum superposition of different addresses $\ket{\psi_{\text{in}}}$ serves as input, and the QRAM produces an entangled state $\ket{\psi_{\text{out}}}$ where each address is correlated with the corresponding memory element \cite{hann2021practicality}:

\begin{equation}
    \ket{\psi_{\text{in}}} = \sum_{i=0}^{N-1} \alpha_i \ket{i}^A \ket{0}^B \xrightarrow{\text{QRAM}} \ket{\psi_{\text{out}}} = \sum_{i=0}^{N-1} \alpha_i \ket{i}^A \ket{x_i}^B~,
\end{equation}

where $N$ is the size of the data vector $\boldsymbol{x}$, and the superscripts $A$ and $B$ represent ``address" and ``bus," respectively.

There are two popular designs for QRAM: fanout QRAM and bucket-brigade QRAM. Their difference primarily lies in how they utilize the address qubit. All QRAM designs consist of quantum routers, which serve to route the input qubit to the right if the router qubit is in state $\ket{1}$ and to the left if the router qubit is in state $\ket{0}~.$ In fanout QRAM, all router qubits are first flipped according to the address qubit. For example, if we want to query the information stored at site $i = 5~,$ the address qubits will be in the computational basis: $\ket{5} = \ket{1} \ket{0} \ket{1}~.$ Then, we need a QRAM with a depth of 3 and flip all the router qubits with a CNOT gate controlled by the address qubits, as illustrated in the figure below.

\begin{figure}[!ht]
\centering
\includegraphics[width=0.625\textwidth]{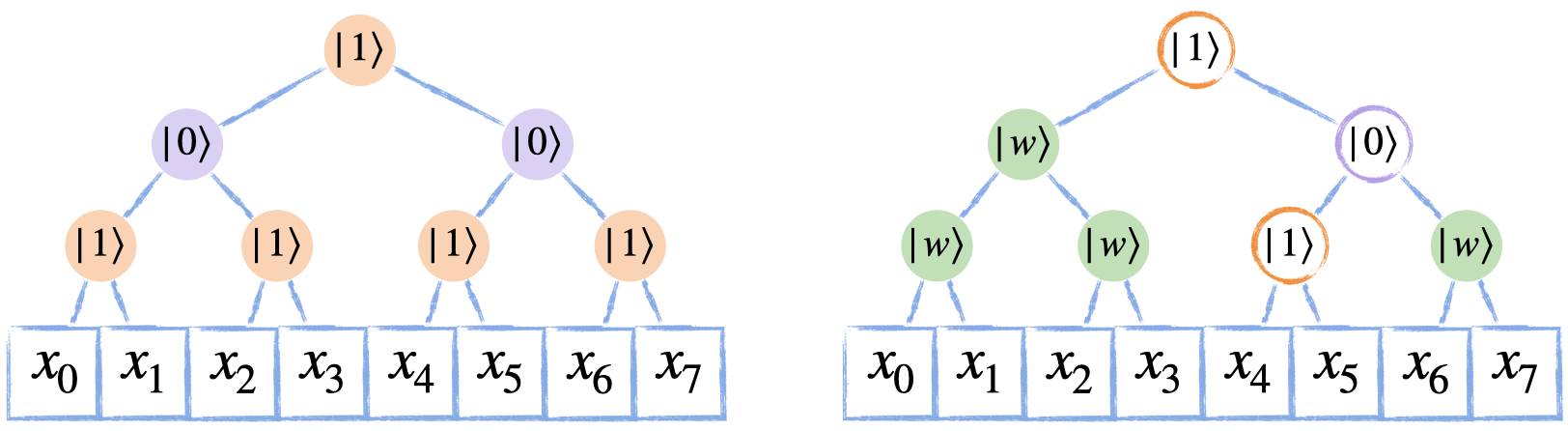}
\caption{The left figure is a picturesque illustration of fanout QRAM \cite{hann2021practicality} to retrieve information stored at $i = 5~.$ On the other hand, the right figure is a picturesque illustration of bucket-brigade QRAM \cite{hann2021practicality} to retrieve information stored at $i = 5~.$ }
\label{Fanout}
\centering
\end{figure}

Then we can route the bus qubit, and since we stored classical information at $x_i$, it is possible for us to copy it to the bus qubit and then route it all the way out before we initialize the QRAM for future use.

The bucket-brigade QRAM is a bit different in the first stage. We don't flip all the router qubits in the layer corresponding to the address qubits. Instead, we route the address qubits into the QRAM one by one, with the router qubits initialized at some state $\ket{w}~.$ Since we route in the address qubits in order, we naturally carve out the path that will lead to the information stored in $x_5~.$

There are debates about whether QRAM is actually practical since it is time-saving but also hardware-consuming. Recently, there are several different QRAM designs that are hardware-efficient. One example is developed by the authors of \cite{Hann_2019}. They made use of circuit quantum acoustodynamic (cQAD) systems \cite{Connell2010quantum,Gustafsson_2014} and constructed quantum gates by applying external drives with different frequencies.

Furthermore, due to its logarithmic depth nature, when QRAM is stored with a large number of data, the speed for querying it will ultimately be bounded by the speed of light. The problem is how many qubits QRAM can store before we hit the boundary set by locality. This problem is addressed in \cite{wang2023fundamental}, where the authors considered locality and found that some 2D or 3D QRAM designs can handle up to $10^{20}$ qubits before hitting the constraint line. This might be already a sufficient number of qubits to perform machine learning training programs. As an result, QRAM might be still reasonable to build in the future. 

Although it looks like that QRAM might be implementable, it is also important that there is no definitive evidence or proof in the hardware especially in the large scale up to date. So whether we can achieve exponential speed ups that some quantum machine learning programs promises might be still unknown and controversial. On the other hand, for some of the QML programs, one can go around QRAM by implementing a quantum subroutine which prepares the quantum states that encodes classical information \cite{biamonte2017quantum}. 

\section{Statistical learning theory} \label{Statistical}

In this section, our aim is to approach a quantum counterpart to statistical learning theory, initiating a systematic framework for delving into the fundamentals of quantum machine learning. A crucial aspect intertwined with statistical learning theory is shadow tomography, addressing a challenge rooted in the distinctive nature of quantum measurements—destructiveness. When concluding a Quantum Machine Learning (QML) program and obtaining a quantum state, extracting classical information becomes a non-trivial endeavor. The challenge arises from the inevitable collapse of the state during measurement, limiting access to the original state. Working with a single copy of an unknown quantum state, denoted as $\rho$, poses an inherent difficulty, as clever strategies cannot approximate a classical description of $\rho$ through direct measurement. The overarching task of reconstructing a description of a $D$-dimensional quantum mixed state $\rho$, given multiple copies of $\rho$, falls under the domain of \textit{quantum state tomography} \cite{aaronson2018shadow}. On the other hand, works such as \cite{Bu_2022,Bu_2023} studied statistical complexity in the context of quantum machine learning, which also closely relates to this topic. 

One significant development inspired by shadow tomography is known as \textit{classical shadow}. As discussed earlier in section \ref{Measurements}, this approach holds promise for addressing the output or downloading problem. Classical shadow infers the output state $\rho$ by performing random measurements, averaging the obtained classical data, and then inverting the results to retrieve the quantum state. This method extensively leverages basics of probability theory to control errors, making it an efficient quantum-to-classical data converter \cite{Elben_2022}. Utilizing classical shadow, one can devise classical machine learning algorithms that exhibit only polynomial differences with full fault-tolerant QML programs, particularly when focusing on average error \cite{huang2021information}. Moreover, this approach holds significant applications in QML for quantum data or quantum simulators, especially when a readout is required at the end.


\subsection{Shadow tomography} \label{Shadow}

\textit{Shadow tomography} was a proposal first came up with by Scott Aaronson couple of years ago \cite{aaronson2018shadow}. In this final section, we will review some key points of it and also a little about its recent developments. 

The foundation of tomography lies in the well-known fact of the destructive nature of measurements in quantum mechanics. When a quantum state or qubit, expressed as $\alpha \ket{0} + \beta \ket{1}$, undergoes a measurement, it collapses into either $\ket{0}$ or $\ket{1}$ with probabilities $\alpha^2$ and $\beta^2$, respectively. Regardless of the outcome, the original state is lost and impossible to recover, a consequence of the No-Cloning Theorem, which prohibits the direct copying of quantum information.





The complexity increases significantly when dealing with $n$ qubits. For a pure state of $n$ qubits, obtaining an approximate description requires dealing with $2^n$ complex numbers. This exponential growth poses a challenge, as it would necessitate exponential classical bits to represent $n$ qubits. Even for a relatively modest system of 1000 qubits, it would require $2^{1000}$ complex numbers for an approximate description. To put this into perspective, this exceeds the total number of atoms in the observable universe, estimated to be around $10^{80}$. This realization, first acknowledged in the 1980s, played a crucial role in the conceptualization of quantum computing by visionaries like Richard Feynman \cite{feynman1996lectures} and others. But again measurement will yield at most $n$ bits. 

The quantum state tomography challenge involves a machine capable of generating $n$-qubit quantum states and producing identical copies of a specified state. Focusing on a mixed quantum state represented by a $D \times D$ density matrix, denoted as $\rho$, the goal is to gain insights into this density matrix. The key question is to determine the minimum number of copies of the state $\rho$ necessary to effectively tackle this challenge. The answer to this question is approximately $\mathcal{O}(D^2)$ copies of $\rho$, which was established in 2016 \cite{odonnell2015efficient}. This implies that the number of required copies scales polynomially with $D$, the dimension of the Hilbert space, which is still exponentially large. For instance, learning about 100 qubits would necessitate around $2^{200}$ copies of the state, an impractical and astronomically large number. Consequently, the feasibility of quantum state tomography becomes increasingly limited as the number of qubits grows, and one of the world records stands at about 10 qubits \cite{Song_2017}, involving millions of measurement settings. This underscores the significant challenges and limitations in quantum state tomography for larger quantum systems.

The concept of \textit{shadow tomography}, introduced by Scott Aaronson \cite{aaronson2018shadow, Aaronson2019Shadow}, is motivated by the following scenario. We are provided with an unknown $D$-dimensional mixed state $\rho$ and a set of observables, $\mathcal{O}_1, \ldots, \mathcal{O}_M$. Assuming these observables can only take two values, such as \textit{yes} or \textit{no}, and considering that these measurements fully characterize the state, we aim to recover the ``shadow" cast by $\rho$ on the measurements $\mathcal{O}_1, \ldots, \mathcal{O}_M$. The term ``shadow tomography" was suggested by Steve Flammia, emphasizing our goal of retrieving the ``shadow" rather than the entire density matrix of $\rho$. The objective is to determine the minimum number of states $\rho$ needed to estimate the expectation value $\operatorname{Pr} [\mathcal{O}_i] = \operatorname{Tr}(\mathcal{O}_i \rho)$ for all these observables.

Our intuition suggests that approximately $\mathcal{O}(D^2)$ copies would be adequate. We can bypass the entire shadow tomography setup and simply obtain a well-approximated version of the state $\rho$. Once we possess the state, accurately determining any expectation values becomes feasible. Furthermore, it seems intuitive that having roughly $\mathcal{O}(M)$ copies of the state should enable straightforward shadow tomography. By conducting measurements on all $M$ observables, we can derive the expectation values. Although the state is lost after each measurement, the availability of numerous copies remains. The latter will be practical if $M$ is small, which means that we only care a small number of measurements or observables. But in many cases, both $D$ and $M$ are enormous. So the crucial question revolves around whether there exist a systematic way of accomplishing shadow tomography can be accomplished with a time complexity of $\operatorname{poly}(\log{D}, \log{M})~.$

In 2018, it was demonstrated in \cite{aaronson2018shadow} that a protocol exists capable of achieving shadow tomography with a time complexity of $\operatorname{poly}(\log{D}, \log{M})$.

\begin{theorem} \label{ShadowTomographyTheorem}
    \textbf{Shadow Tomography Theorem}
    
    Shadow tomography is solvable using only $\tilde{\mathcal{O}}\left( \log^4{M} \cdot \log{D} / \varepsilon^4 \right)$ copies of the state $\rho~,$ where the $\tilde{\mathcal{O}}$ hides a higher order factor. The procedure is fully explicit. 
\end{theorem}

In simpler terms, this protocol estimates the acceptance probability of all measurements, denoted as $\mathcal{O}_1, \ldots, \mathcal{O}_M$, with an additive error of $\pm \varepsilon$, utilizing only approximately $\tilde{\mathcal{O}}\left( \log^4{M} \cdot \log{D}/\varepsilon^4 \right)$ copies of $\rho$. While it's acknowledged that this is not the most efficient method and there is room for improvement, this protocol marked a significant milestone by demonstrating the theoretical feasibility of shadow tomography.

The implications of Theorem \ref{ShadowTomographyTheorem} include the following:

\begin{enumerate}
    \item For an $n$-qubit state $\ket{\psi}$, by measuring approximately $n^{\mathcal{O}(1)}$ copies of $\ket{\psi}$, one can understand its behavior on \textbf{every} quantum circuit with at most $p(n)$ gates, for any fixed polynomial $p$.
    \item Given any $n^{\mathcal{O}(1)}$-qubit quantum program $\rho$, having $n^{\mathcal{O}(1)}$ copies of $\rho$ allows us to estimate its acceptance probability on \textbf{every} $n$-bit input $x$.
\end{enumerate} 

To comprehend the workings of shadow tomography, a key concept is ``gentle measurement" \cite{aaronson2019gentle}. If a measurement of a mixed state $\rho$ yields a certain outcome with a probability greater than or equal to $1-\varepsilon$, then after the measurement, there exists a state $\rho'$ that is $\sqrt{\varepsilon}$-close to $\rho$ in trace distance. In simpler terms, if a state is close to an eigenstate of some observable, it is highly likely that the state will essentially collapse into that eigenstate, which isn't much different from the original state before the measurement occurred. An even stronger statement holds: applying $M$ such measurements in succession, each accepting with a probability greater than or equal to $1-2M\sqrt{\varepsilon}$, ensures that the error grows linearly. Thus, if a sequence of measurements is applied, and each one does not significantly damage $\rho$, the cumulative effect of the entire session (assuming not too many measurements, i.e., small $M$) will also not significantly damage $\rho$.

The concept of gentle measurement is often paired with \textit{amplification}. If one has many identical copies of a state $\rho$, and the measurement on one copy yields a certain outcome with a $90$ percent probability, then with $k$ copies of the state, an \textit{amplified measurement} can be applied. This involves coherently measuring $k$ copies of $\rho$ and taking a \textbf{majority} to reduce the error to $1/\exp{k}$. Therefore, one can use amplification initially to reduce $\varepsilon$ or the probability of a bad measurement to an exponentially small value with respect to $k$, the number of copies measured. Subsequently, gentle measurement can be employed, ensuring that the damage to the state $\rho$ each time a measurement is applied is only of an exponentially small amount.

One obvious question might be why gentle measurement does not solve our problem of shadow tomography immediately. Gentle measurement only provides a solution under a special case known as a \textbf{promise gap}. The gentle measurement will give us an answer under the following circumstance: given measurements $\mathcal{O}_1, \ldots, \mathcal{O}_M$ and real numbers $C_1, \ldots, C_M$, and being promised that $\operatorname{Pr}[\mathcal{O}_i \text{ accepts } \rho]$ is either at least $C_i$ or at most $C_i - \varepsilon$ for each $i$, we can decide for each $i$ using only $\mathcal{O} \left( \log{M}/\varepsilon^2 \right)$ copies of $\rho$. This provides a good zone for gentle measurements to figure out which side of the two outcome measurements our state is close to, allowing gentle measurement to work. The problem arises when we are in the intermediate zone: $C_i - \varepsilon \geq \operatorname{Pr}[\mathcal{O}_i \text{ accepts } \rho] \geq C_i$. In this case, we will not be able to use this trick and accomplish shadow tomography. In other words, when there's no promise, we can never rule out that we're on the knife-edge between acceptance and rejection, making measuring $\rho$ dangerous, since measurement there will not be gentle and destroy our state. And because our objective limits our copies of $\rho$ to only $\log{M}$, we can't afford to destroy that many copies of the state.

The whole point of shadow tomography is to address this problem \cite{aaronson2018shadow,Aaronson2019Shadow}. The actual proof of Theorem \ref{ShadowTomographyTheorem} involves a combination of several tricks. The first one was intended to solve a different yet similar problem. Imagine Alice, who knows the state $\rho$, needs to send its information or description to Bob with respect to its behavior on measurements of observables $\mathcal{O}_1, \ldots, \mathcal{O}_M$, which we call $E_1, \ldots, E_M$. In contrast to shadow tomography, Alice does know the state at the beginning, but she has to send the information about the measurements in much fewer bits.

The following program will solve this problem, resembling a machine learning program. Initially, since Bob knows absolutely nothing, he will take a guess, and let's set it to the \textit{maximally entangled state}: $\rho_0$. Then Alice helps Bob improve by repeatedly telling him a measurement with respect to an observable $\mathcal{O}_{i(t)}$ on which Bob's current guess $\rho_{t-1}$ badly fails. Bob lets $\rho_t$ be the state obtained by starting from $\rho_{t-1}$, then performing the measurement of $\mathcal{O}_{i(t)}$ and \textbf{postselecting} on getting the right outcome. After a few rounds, Bob has improved his state. To clarify, Bob is not trying to learn $\rho$ at all but only $\rho$'s behavior. He might end up with a state that is very far away in terms of trace distance from $\rho$ yet still improves with respect to all the measurements that we care about. So the extreme situation is that Bob would guess randomly and obtain a very different state, yet this state has nothing to do with the measurements to improve upon. Then Bob is done on day one. The key point is for boosting weights type reasons; this process must converge after $T = \mathcal{O}(\log{D})$ iterations, at some state $\rho_{T}$ that behaves like $\rho$ in terms of $E_1, \ldots, E_M$, even if it's far from $\rho$ in trace distance.

Can we use this postselection learning protocol to solve the problem of shadow tomography? Unfortunately, it's still not enough by itself. Here, we relied on Alice to know the state $\rho$, and to know which measurement to be used that is best fit for postselection and updating Bob's guess. However, in the shadow tomography situation, there is no Alice. There is no one who knows the state and can tell you useful measurements to condition on. Rather, you just have these copies of a $\rho$, and we have to figure it out for ourselves.

Another protocol to be combined with the previous one is called the \textbf{Quantum OR Bound} \cite{Harrow_2017}.

\begin{theorem}
    \textbf{Quantum OR Bound}

    Let $\rho$ be an unknown mixed state, and let $E_1~, \ldots~, E_M$ be known 2-outcome measurements. Suppose we are promised that either

    \begin{enumerate}
        \item \textbf{there exists} an $i$ such that $\operatorname{Pr}[E_i \text{ accepts } \rho] \geq C~;$ or else
        \item $\operatorname{Pr}[E_i \text{ accepts } \rho] \leq C - \varepsilon~,$~ \textbf{for all} $i \in [M]~.$
    \end{enumerate}
    Then we can decide which, with high probability, given only $\mathcal{O} \left( \log{M} / \varepsilon^2 \right)$ copies of $\rho~.$
\end{theorem}

So what's new about shadow tomography is the combination of two protocols, \textbf{postselection learning} and the \textbf{Quantum OR Bound}. Scott Aaronson \cite{aaronson2018shadow} found that you can use the Quantum OR Bound protocol as a subroutine to repeatedly search for informative measurements. Again, you start with the maximally mixed state and use the quantum OR bound repeatedly to search for measurements that can best improve your guess, round by round, to shape it into the one that has a similar outcome in terms of measurements with the target state $\rho$. Putting everything together, one will end up with Theorem \ref{ShadowTomographyTheorem}. 

The result in Theorem \ref{ShadowTomographyTheorem} is not the upper or lower bound; it simply shows that shadow tomography can be done in $\log{M}$ steps. Nowadays, the best lower bound people can prove is $\mathcal{O}(\log{M}/\varepsilon^2)$, which holds even if we are trying to learn a classical distribution \cite{aaronson2019gentle}. So it is still unclear whether we need quantum Hilbert dimension $D$ in Theorem \ref{ShadowTomographyTheorem}. In \cite{Aaronson2019Shadow}, the authors were able to improve upon the $\log{M}$ dependence using a brand-new connection between gentle measurement and differential privacy. The sample complexity they have achieved is $    k = \mathcal{O} \left( \log^2{M} \cdot \log^2{D} / \varepsilon^8 \right)~.$

Inspired by shadow tomography, \cite{huang2021information} introduced the classical shadow protocol, as discussed in section \ref{Measurements}. Thus, quantum machine learning is a rapidly developing field that encompasses various topics in physics and computer science. Although sub-fields may appear distinct, they all fall under the umbrella of quantum machine learning, influencing and potentially finding applications in one another. As a result, as Scott Aaronson states at the end of his paper \cite{aaronson2015read}, despite decades of research in quantum computing, researchers still marvel at the fact that the laws of quantum physics enable us to solve classical problems significantly faster than today's computers seem capable of. Therefore, it should perhaps not surprise us that, in machine learning and elsewhere, unlocking the full potential of quantum speedups requires significant effort.

\subsection{Classical shadow formalism and random measurements} \label{CAlgorithms}

Drawing inspiration from the principles of shadow tomography, the classical shadow formalism has emerged as an intriguing topic with the potential to serve as an efficient quantum-classical information converter. A series of works, such as \cite{Huang_2020, huang2021information, Elben_2022}, have contributed to establishing a rigorous framework for reasoning about the randomized measurement paradigm, a key component of classical shadow. These works not only delve into the conceptual aspects but also provide methods to derive error bounds for quantum information extraction based on randomized measurements, leveraging probability theory.

The challenges associated with the quantum-classical information interface, particularly in the context of NISQ algorithms (discussed in sections \ref{VQA} and \ref{QNTK}), as well as in the FTQC era for quantum machine learning algorithms (explored in section \ref{Carleman}), highlight the necessity of efficient readout mechanisms. Researchers are actively addressing the need for reliable ways to input and extract information from quantum systems to classical ones, as the success of quantum machine learning programs relies on overcoming these challenges. If the process of loading classical data into a quantum computer or transforming quantum data into classical form takes an significant amount of time, the promised speed-ups of quantum algorithms become impractical. Therefore, not only is it crucial to find methods for transforming classical data into quantum form and vice versa, but the efficiency of these processes, including achieving significant speed-up, is paramount, given the exponential growth of Hilbert space degrees of freedom. 

The most direct and comprehensive solution to the input problem is Quantum Random Access Memory (QRAM), as detailed in section \ref{QRAM}. Additionally, alternative techniques like pruning, discussed in section \ref{Carleman}, offer potential solutions to circumvent the pressing requirement for QRAM. Classical shadow, see also section \ref{Measurements}, on the other hand, is focused on addressing the output problem. In this scenario, we possess a quantum state that encapsulates information trained by our Quantum Machine Learning (QML) program, and the goal is to decode this quantum state into classical information for straightforward interpretation. 

To present this challenge in a broader context, the task is to efficiently estimate the expectation values of multiple structured observables. Specifically, the aim is to estimate the expectation value of various individual terms in Hamiltonians, where these terms share a well-defined structure, such as being Pauli operators. The proposed strategy for addressing this challenge involves diverting attention from the intricate details of the terms and instead uniformly sampling Pauli measurements. In simpler terms, the goal is to estimate the expectation values of Pauli observables, with the selection of the Pauli basis for each qubit being determined randomly during the measurement process \cite{Elben_2022}.

\begin{figure}[!ht]
\centering
\includegraphics[width=0.7\textwidth]{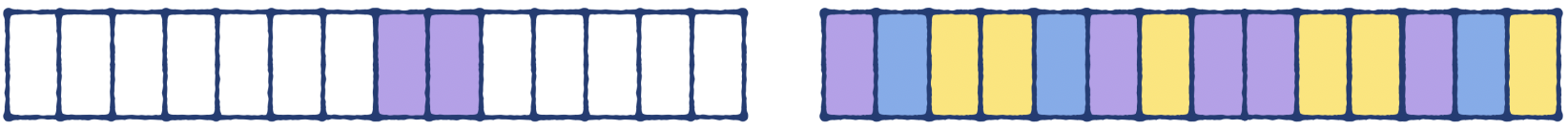}
\caption{A picturesque illustration of the infinite monkey theorem applied to randomized measurements.}
\label{InfiniteMonkey}
\centering
\end{figure}

This concept can be grasped intuitively through the lens of the infinite monkey theorem \cite{evans2019scalable}. The theorem posits that a monkey randomly hitting keys on a typewriter for an infinite amount of time will almost certainly type any given text, including the complete works of William Shakespeare, and will likely produce every possible finite text an infinite number of times. Let's apply this idea to predicting an arbitrary low-weight Pauli observable, represented by the purple blocks in Fig.~\ref{InfiniteMonkey}. The white blocks signify identity matrices, which are not of primary concern. The crucial aspect is the presence of three specific Pauli operators adjacent to each other in Fig.~\ref{InfiniteMonkey}. Now, if we randomly assign Pauli measurements to each qubit, we are free to choose any Pauli measurements for the white blocks on the left side of Fig.~\ref{InfiniteMonkey}. However, for the purple blocks, we must use the same Pauli measurements as on the left, or else we won't gain meaningful information about the statistics. To achieve this, random sampling is employed. Importantly, the performance is not adversely affected, as the probability of obtaining such a low-weight Pauli string is exponentially small, but this exponentiation is in the weight of the Pauli, not in the number of qubits \cite{evans2019scalable}. Additionally, if single-shot measurements are considered, there will be an oversampling factor.

The crucial aspect is that, since everything is done randomly, an effective union bound can be applied to estimate numerous low-weight Pauli observables. The union bound scales logarithmically with the number of terms. Consequently, if one randomly samples each Pauli measurement, the expected total number of measurements required to predict $L$ low-weight observables with accuracy $\epsilon$ each is only logarithmic in $L$, thanks to the union bound \cite{Elben_2022}.

\begin{figure}[!ht]
\centering
\includegraphics[width=0.7\textwidth]{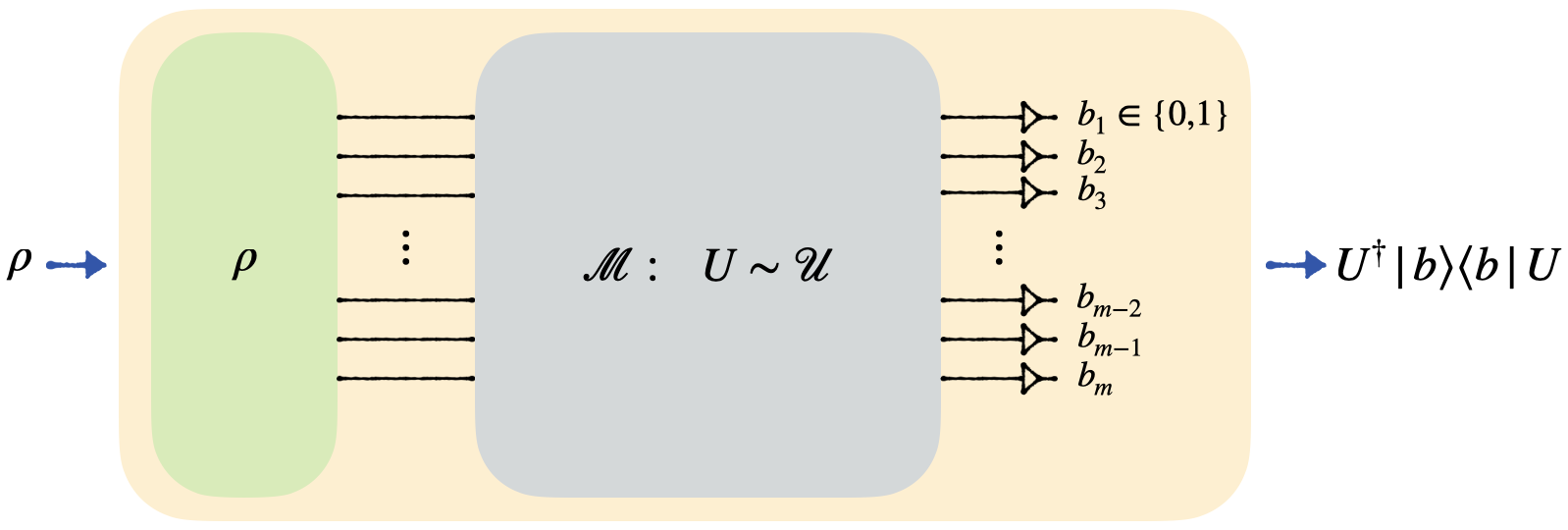}
\caption{A quantum to classical channel.}
\label{QtoC}
\centering
\end{figure}

To formally understand the classical shadow formalism, first we construct a quantum channel which transforms the quantum data into the classical data. We start with an unknown $m$ qubit quantum state $\rho$ and do a certain random unitary rotation: $U~,$ for example an independent single qubit Clifford rotation. Then we do a computational basis measurement on all of them. Since we did measurement at last, so the output of this channel is actually, honest to god, classical information. This, in Fig.~\ref{QtoC}, means that $U^{\dagger} \ketbra{b}{b} U$ although written in quantum language, is classical data. Hence, altogether, what this channel does is to transform the quantum state we got from the quantum computer and mapped it into classical information which we will do further computation. 

The subsequent step involves averaging the random unitary sampling (through an integral over $U$), depending on the chosen ensemble, and also averaging over the computational basis outcomes (by summing over $b$) as per Born's rule. If the ensemble is well-behaved, such as Haar random, then the result after the averaging step can be efficiently computed. Moreover, since the result is classical information rather than quantum, it can be easily inverted to obtain the state $\rho$. Another example may be found in \cite{Bu_2024}, where the authors consider classical shadows with unitary ensembles that are invariant under multiplication by Pauli operators. Consequently, the state $\rho$ is obtained by performing random measurements to acquire classical data, taking average, and then inverting the result to retrieve the state. The lingering question pertains to the convergence speed of this program to the desired expectation value. Drawing inspiration from shadow tomography, the authors of \cite{Huang2020Predeicting} demonstrated that the number of required snapshots or single measurements in this program scales logarithmically with the number of terms needed to predict: $N \gtrsim \log{(L)} \max_j ||\mathcal{O}_j||_{\text{shadow}}^2/\epsilon^2$, where $\mathcal{O}_j$ is the observable to be measured, and $\epsilon$ denotes the desired accuracy.

In 2021, the authors of \cite{Struchalin2021Experimental} conducted an experimental implementation of classical shadow, demonstrating its superiority over maximum likelihood in the low sampling regime. This framework's applicability extends to the prediction of observable polynomials, as discussed in \cite{Elben_2022}. The strategy involves expressing polynomials as linear combinations of tensor products and substituting the tensor products with independent shadows, enhancing efficiency in probing entanglements in many-body systems. Notably, this method exhibits robustness in the presence of noise, leveraging random unitaries as detailed in \cite{Chen_2021}. The authors of \cite{Koh_2022} propose an error-mitigated classical shadow estimation scheme. This result complements \cite{Chen_2021}. \cite{wu2023errormitigated} introduced an error-mitigated fermionic classical shadow scheme. Additionally, \cite{Huang2021Efficient} explores techniques for derandomizing the procedure. In classical computer science, there exists a standard approach to transform a probabilistic argument into a deterministic strategy, maintaining comparable effectiveness without reliance on randomness.

\subsubsection{Machine learning with classical shadow} \label{MLClassicalShadow}

In summary, classical shadows serve as an efficient quantum-to-classical converter, enabling the extraction of classical information from quantum states. This information can be used to predict observable expectation values. When considering the Variational Quantum Eigensolver (VQE) in a classical context, it resembles a kernel method. The process involves initiating with parameters that define the circuit, creating a map in a high-dimensional space. These parameters are then used to map to a state generated by the circuit, constructing an objective function with the Hamiltonian to acquire classical information. Moreover, classical shadows have demonstrated significant performance gains for VQE \cite{Elben_2022}, prompting exploration into their potential implementation for more sophisticated machine learning programs.

Consider the problem of predicting ground state properties, where we assume a family of parameterized Hamiltonians, denoted as $H(\boldsymbol{x})$, with $\boldsymbol{x}$ being a classical input characterizing the Hamiltonian (e.g., Heisenberg Hamiltonian with $\boldsymbol{x}$ representing couplings). The goal is to learn how to predict ground state properties, such as observable expectation values, for ground states associated with Hamiltonians that were not part of the training data.

Classically, this problem is challenging. However, leveraging quantum training data and classical shadow measurements simplifies the task. Here's how the process unfolds:

\begin{enumerate}
    \item \textbf{Learning Phase}:
    \newline
   - Random input parameters are sampled, and the associated ground states are prepared in the laboratory.
    \newline
   - Classical shadow measurements are performed on these ground states for randomly chosen single qubits.
    \newline
   - The training data comprises classical inputs ($\boldsymbol{x}$) associated with classical shadow single-shot measurements of the corresponding ground states.
   \newline
   - The goal is to train a classical kernel method using these classical shadows.

    \item \textbf{Prediction Phase}:
    \newline
   - In this phase, a Fourier-type kernel function is employed to convolve all the classical shadows observed during the learning phase.
   \newline
   - This convolution provides an approximation of the ground state for Hamiltonians not directly observed.
\end{enumerate}

The central concept involves using classical shadows as a connection between quantum training data and classical predictions. This approach facilitates the learning of ground state properties for Hamiltonians that were not part of the initial training set. The goal is to establish an efficient representation for future Hamiltonians, enabling the computation of observable expectation values and making predictions. This program has been demonstrated to be effective. In \cite{Huang_2022}, the authors established that for a small average case prediction error (averaged over all possible inputs), a scenario common in machine learning, the required amount of training data is not extensive. Instead, it scales quasi-polynomially in the number of inputs, not the number of qubits. 

In the study conducted by \cite{huang2021information}, the authors delved deeper into the efficacy of this method. They utilized a quantum device to generate data, employed classical shadows to effectively utilize the quantum data, and then applied classical machine learning algorithms for training. They compared this approach (classical machine learning with quantum data leading to classical shadows) to a hypothetical full-scale quantum machine learning protocol. The latter runs on a quantum computer and has coherent access to training data through QRAM. This is the most powerful program that we can envision in the future of QML. The results indicated that, when focusing on average error, there is no significant separation between quantum and classical training data size in terms of sample complexity. However, for worst-case prediction error, \cite{huang2021information} presented concrete examples demonstrating an exponential separation in complexity. In the research presented in \cite{huang2022quantum}, the authors explore analogous approaches to learning. They investigate how classical and quantum agents, given their distinct methods of receiving, processing, and storing information, differ in efficiency when aiming to characterize a physical system or state $\rho$. The study demonstrates that quantum agents can achieve exponential advantages in experiments, particularly in terms of the number of queries required to obtain $\rho$.

\subsection{Quantum machine learning for quantum data and quantum simulators} \label{QuantumData}

The most immediate application of quantum machine learning lies in handling quantum data \cite{biamonte2017quantum}, i.e., the states produced by quantum systems and processes. Traditional quantum machine learning algorithms, as discussed earlier, identify patterns in classical data by transforming it into quantum states and manipulating those states using basic quantum linear algebra operations. These same algorithms can be directly employed on quantum states of light and matter to uncover inherent features and patterns. The quantum methods of analysis often prove to be more efficient and insightful than classical approaches when dealing with data obtained from quantum systems.

For instance, in the case of multiple copies of a system represented by an $N \times N$ density matrix, quantum principal component analysis can determine eigenvalues and reveal corresponding eigenvectors in a time complexity of $\mathcal{O}\left(\left(\log_2{N} \right)^2 \right)$. This stands in contrast to the $\mathcal{O}(N^2)$ measurements required for classical devices to perform tomography on the density matrix and the $\mathcal{O}(N^2$) operations needed for classical PCA. This quantum analysis of quantum data can be efficiently conducted on the relatively modest-scale quantum computers expected to be available in the coming years. As mentioned before, \cite{huang2022quantum} also demonstrated the quantum advantages of learning a physical state $\rho$ through a quantum agent for example quantum sensor or simulator discussed above. 


Quantum simulators serve as powerful tools for analyzing quantum dynamics. These are essentially ``quantum analog computers" capable of emulating the dynamics of specific quantum systems. Quantum simulators can be specialized devices designed for simulating particular classes of quantum systems or general-purpose quantum computers. The approach involves connecting a trusted quantum simulator to an unknown system, adjusting the simulator's model to counteract the unknown dynamics, and efficiently learning the dynamics of the unknown system using approximate Bayesian inference \cite{biamonte2017quantum, Granade_2012, Wiebe_2014, Wiebe_2015}. This significantly reduces the number of measurements required for simulation, offering an exponential improvement.

Notably, the universal quantum emulator algorithm \cite{marvian2016universal} enables the reconstruction of quantum dynamics, and the quantum Boltzmann training algorithm \cite{biamonte2017quantum} allows for the reconstruction of states with a time complexity logarithmic in the dimension of the Hilbert space. This is significantly faster than reconstructing dynamics through classical tomography.

While using a quantum computer to characterize a quantum system \cite{Wiebe_2014, Wiebe_2015} or input states for quantum PCA poses technical challenges, as it doesn't require QRAM, it holds promise for near-term applications of quantum machine learning \cite{Wiebe_2014, Wiebe_2015,biamonte2017quantum}, offering the potential for significant speedups in device characterization.

\section{Conclusion and discussion}

This review highlights that both current NISQ-era quantum devices and future fully fault-tolerant quantum computers (FTQC) hold significant promise for applications in machine learning and data analysis \cite{bharti2021noisy,biamonte2017quantum,liu2022representation,liu2023towards}. Over the past decade, quantum computing has witnessed notable advancements in applications, experimental validations, and theoretical findings. The quantum computation field, especially in NISQ applications, has seen an almost exponential growth in the number of research papers. Various factors contribute to this trend, including substantial enhancements in quantum hardware \cite{bharti2021noisy, huang2022quantum}. Given that quantum computing is a relatively young scientific discipline, there is ample opportunity for groundbreaking research and discoveries. The presence of theoretical, practical, and experimental challenges, many of which are addressed in this review, further underscores the motivation for an open-source approach in the field.

As mentioned in \cite{bharti2021noisy}, we expect experimental pursuit in the NISQ era would focus on the design of quantum hardware with a larger number of qubits, and gates with lower error rates capable of executing deeper circuits. Along the way, one of the goals is to demonstrate quantum advantage for practical use cases. If the NISQ paradigm is not powerful enough to exhibit any quantum advantage, theoretical pursuits would be required to understand its limitations. The prime direction of the NISQ and near-term era is to engineer the best possible solution with the limited quantum resources available. The tools and techniques invented during this period could be valuable in the fault- tolerant era as well. To conduct a successful demonstration of quantum advantage, the right blend of the two crucial components: hardware and algorithms designs, is required \cite{bharti2021noisy}. First, hardware development is the key. The design of quantum computers with more qubits, lesser error rates, longer coherence times, and more connectivity between the qubits will be one of the top priorities in the NISQ era. Intensive research in new qubits developments, quantum optimal control and material discovery will be indispensable for both universal programmable quantum computers or special purpose ones. Secondly, to harness the potential of noisy but powerful quantum devices, we expect breakthroughs on the algorithm frontier. Algorithms with realistic assumptions, as the ones mentioned in \cite{bharti2021noisy}, regarding device capabilities will be favored. To lessen the effect of noise, progress towards the design of error suppression, mitigation and correction methods is expected. 

While QML has been proposed as a potential avenue for achieving scientific value in the near term using NISQ devices, questions arise regarding its applicability in the future \cite{cerezo2022challenges}. Researchers envision two distinct post-NISQ eras. The first, termed ``partial error corrected," assumes that quantum computers will possess a sufficient number of physical qubits (a few hundred) and low error rates, allowing for a small number of fully error-corrected logical qubits. In this era, users can allocate qubits between error-corrected and non-error-corrected subsets. The subsequent era, labeled the ``fault-tolerant era," will emerge when quantum hardware features a large number of error-corrected qubits. In this era, algorithms like those discussed in section \ref{Carleman} could notably enhance the scalability and sustainability of classical large-scale machine learning models. Notably, works such as \cite{Liu2021Efficient} offer solid theoretical assurances and intersections with cutting-edge classical machine learning research. This approach diverges from the variational quantum algorithms mindset, exemplified by \cite{liu2022representation}, by aiming to enhance classical machine learning through a key quantum step that acts as a bottleneck for classical training. Developments in shadow tomography and related fields \cite{aaronson2018shadow, Elben_2022, huang2021information} demonstrate the potential for systematically analyzing QML and gaining a deeper understanding of how quantum speedup is achieved. As demonstrated experimentally in \cite{huang2022quantum}, with quantum technology such as quantum sensor, quantum memory, and quantum computer, our ability to learn about the physical world might be exponentially improved. Hence, there is confidence that in the future, researchers will have a much more robust understanding of how quantum computing achieves speedup and can develop more powerful applications for machine learning problems. 



\section*{Acknowledgement}
We thank Hsin-Yuan Huang and Kanav Setia for helpful discussions. JL is supported in part by International Business Machines (IBM) Quantum through the Chicago Quantum Exchange, and the Pritzker School of Molecular Engineering at the University of Chicago through AFOSR MURI (FA9550-21-1-0209).

\newcommand{\newblock}{}
\bibliography{QML.bib}

\begin{thebibliography}{139}
\providecommand{\natexlab}[1]{#1}
\providecommand{\url}[1]{\texttt{#1}}
\expandafter\ifx\csname urlstyle\endcsname\relax
  \providecommand{\doi}[1]{doi: #1}\else
  \providecommand{\doi}{doi: \begingroup \urlstyle{rm}\Url}\fi

\bibitem[Aaronson(2015)]{aaronson2015read}
S.~Aaronson.
\newblock Read the fine print.
\newblock \emph{Nature Physics}, 11\penalty0 (4):\penalty0 291--293, 2015.

\bibitem[Aaronson(2018)]{aaronson2018shadow}
S.~Aaronson.
\newblock Shadow tomography of quantum states.
\newblock In \emph{Proceedings of the 50th annual ACM SIGACT symposium on theory of computing}, pages 325--338, 2018.

\bibitem[Aaronson(2020)]{Aaronson2019Shadow}
S.~Aaronson.
\newblock Shadow tomography of quantum states.
\newblock \emph{SIAM Journal on Computing}, 49\penalty0 (5):\penalty0 STOC18--368--STOC18--394, 2020.
\newblock \doi{10.1137/18M120275X}.
\newblock URL \url{https://doi.org/10.1137/18M120275X}.

\bibitem[Aaronson and Rothblum(2019)]{aaronson2019gentle}
S.~Aaronson and G.~N. Rothblum.
\newblock Gentle measurement of quantum states and differential privacy, 2019.

\bibitem[Amaro et~al.(2022)Amaro, Modica, Rosenkranz, Fiorentini, Benedetti, and Lubasch]{Amaro_2022}
D.~Amaro, C.~Modica, M.~Rosenkranz, M.~Fiorentini, M.~Benedetti, and M.~Lubasch.
\newblock Filtering variational quantum algorithms for combinatorial optimization.
\newblock \emph{Quantum Science and Technology}, 7\penalty0 (1):\penalty0 015021, Jan. 2022.
\newblock ISSN 2058-9565.
\newblock \doi{10.1088/2058-9565/ac3e54}.
\newblock URL \url{http://dx.doi.org/10.1088/2058-9565/ac3e54}.

\bibitem[Anand et~al.(2021)Anand, Degroote, and Aspuru-Guzik]{Anand_2021Natual}
A.~Anand, M.~Degroote, and A.~Aspuru-Guzik.
\newblock Natural evolutionary strategies for variational quantum computation.
\newblock \emph{Machine Learning: Science and Technology}, 2\penalty0 (4):\penalty0 045012, jul 2021.
\newblock \doi{10.1088/2632-2153/abf3ac}.
\newblock URL \url{https://doi.org/10.1088%2F2632-2153%2Fabf3ac}.

\bibitem[Andreassen et~al.(2019)Andreassen, Feige, Frye, and Schwartz]{andreassen2019junipr}
A.~Andreassen, I.~Feige, C.~Frye, and M.~D. Schwartz.
\newblock Junipr: a framework for unsupervised machine learning in particle physics.
\newblock \emph{The European Physical Journal C}, 79:\penalty0 1--24, 2019.

\bibitem[Arrasmith et~al.(2021)Arrasmith, Cerezo, Czarnik, Cincio, and Coles]{Arrasmith_2021}
A.~Arrasmith, M.~Cerezo, P.~Czarnik, L.~Cincio, and P.~J. Coles.
\newblock Effect of barren plateaus on gradient-free optimization.
\newblock \emph{Quantum}, 5:\penalty0 558, Oct. 2021.
\newblock ISSN 2521-327X.
\newblock \doi{10.22331/q-2021-10-05-558}.
\newblock URL \url{http://dx.doi.org/10.22331/q-2021-10-05-558}.

\bibitem[Arrasmith et~al.(2022)Arrasmith, Holmes, Cerezo, and Coles]{Arrasmith_2022}
A.~Arrasmith, Z.~Holmes, M.~Cerezo, and P.~J. Coles.
\newblock Equivalence of quantum barren plateaus to cost concentration and narrow gorges.
\newblock \emph{Quantum Science and Technology}, 7\penalty0 (4):\penalty0 045015, Aug. 2022.
\newblock ISSN 2058-9565.
\newblock \doi{10.1088/2058-9565/ac7d06}.
\newblock URL \url{http://dx.doi.org/10.1088/2058-9565/ac7d06}.

\bibitem[Arunachalam et~al.(2015)Arunachalam, Gheorghiu, Jochym-O’Connor, Mosca, and Srinivasan]{Arunachalam_2015}
S.~Arunachalam, V.~Gheorghiu, T.~Jochym-O’Connor, M.~Mosca, and P.~V. Srinivasan.
\newblock On the robustness of bucket brigade quantum ram.
\newblock \emph{New Journal of Physics}, 17\penalty0 (12):\penalty0 123010, Dec. 2015.
\newblock ISSN 1367-2630.
\newblock \doi{10.1088/1367-2630/17/12/123010}.
\newblock URL \url{http://dx.doi.org/10.1088/1367-2630/17/12/123010}.

\bibitem[au2 et~al.(2023)au2, Zaman, and Wong]{morrell2023stepbystep}
H.~J. M.~J. au2, A.~Zaman, and H.~Y. Wong.
\newblock Step-by-step hhl algorithm walkthrough to enhance the understanding of critical quantum computing concepts, 2023.

\bibitem[Barak and Marwaha(2021)]{barak2021classical}
B.~Barak and K.~Marwaha.
\newblock Classical algorithms and quantum limitations for maximum cut on high-girth graphs.
\newblock \emph{arXiv preprint arXiv:2106.05900}, 2021.

\bibitem[Barkoutsos et~al.(2020)Barkoutsos, Nannicini, Robert, Tavernelli, and Woerner]{barkoutsos2020improving}
P.~K. Barkoutsos, G.~Nannicini, A.~Robert, I.~Tavernelli, and S.~Woerner.
\newblock Improving variational quantum optimization using cvar.
\newblock \emph{Quantum}, 4:\penalty0 256, 2020.

\bibitem[Becker et~al.(2023)Becker, Englund, and Stiller]{becker2023optoacoustic}
S.~Becker, D.~Englund, and B.~Stiller.
\newblock An optoacoustic field-programmable perceptron for recurrent neural networks.
\newblock \emph{arXiv preprint arXiv:2309.01543}, 2023.

\bibitem[Benedetti et~al.(2019)Benedetti, Garcia-Pintos, Perdomo, Leyton-Ortega, Nam, and Perdomo-Ortiz]{benedetti2019generative}
M.~Benedetti, D.~Garcia-Pintos, O.~Perdomo, V.~Leyton-Ortega, Y.~Nam, and A.~Perdomo-Ortiz.
\newblock A generative modeling approach for benchmarking and training shallow quantum circuits.
\newblock \emph{npj Quantum Information}, 5\penalty0 (1):\penalty0 45, 2019.

\bibitem[Benedetti et~al.(2021)Benedetti, Fiorentini, and Lubasch]{Benedetti_2021}
M.~Benedetti, M.~Fiorentini, and M.~Lubasch.
\newblock Hardware-efficient variational quantum algorithms for time evolution.
\newblock \emph{Physical Review Research}, 3\penalty0 (3), July 2021.
\newblock ISSN 2643-1564.
\newblock \doi{10.1103/physrevresearch.3.033083}.
\newblock URL \url{http://dx.doi.org/10.1103/PhysRevResearch.3.033083}.

\bibitem[Bharti et~al.(2021)Bharti, Cervera-Lierta, Kyaw, Haug, Alperin-Lea, Anand, Degroote, Heimonen, Kottmann, Menke, et~al.]{bharti2021noisy}
K.~Bharti, A.~Cervera-Lierta, T.~H. Kyaw, T.~Haug, S.~Alperin-Lea, A.~Anand, M.~Degroote, H.~Heimonen, J.~S. Kottmann, T.~Menke, et~al.
\newblock Noisy intermediate-scale quantum (nisq) algorithms (2021).
\newblock \emph{arXiv preprint arXiv:2101.08448}, 2021.

\bibitem[Biamonte et~al.(2017)Biamonte, Wittek, Pancotti, Rebentrost, Wiebe, and Lloyd]{biamonte2017quantum}
J.~Biamonte, P.~Wittek, N.~Pancotti, P.~Rebentrost, N.~Wiebe, and S.~Lloyd.
\newblock Quantum machine learning.
\newblock \emph{Nature}, 549\penalty0 (7671):\penalty0 195--202, 2017.

\bibitem[Bilkis et~al.(2023)Bilkis, Cerezo, Verdon, Coles, and Cincio]{Bilkis_2023}
M.~Bilkis, M.~Cerezo, G.~Verdon, P.~J. Coles, and L.~Cincio.
\newblock A semi-agnostic ansatz with variable structure for variational quantum algorithms.
\newblock \emph{Quantum Machine Intelligence}, 5\penalty0 (2), 2023.
\newblock ISSN 2524-4914.
\newblock \doi{10.1007/s42484-023-00132-1}.
\newblock URL \url{http://dx.doi.org/10.1007/s42484-023-00132-1}.

\bibitem[Bittel and Kliesch(2021)]{Bittel2021Training}
L.~Bittel and M.~Kliesch.
\newblock Training variational quantum algorithms is np-hard.
\newblock \emph{Phys. Rev. Lett.}, 127:\penalty0 120502, Sep 2021.
\newblock \doi{10.1103/PhysRevLett.127.120502}.
\newblock URL \url{https://link.aps.org/doi/10.1103/PhysRevLett.127.120502}.

\bibitem[Bu et~al.(2022)Bu, Koh, Li, Luo, and Zhang]{Bu_2022}
K.~Bu, D.~E. Koh, L.~Li, Q.~Luo, and Y.~Zhang.
\newblock Statistical complexity of quantum circuits.
\newblock \emph{Physical Review A}, 105\penalty0 (6), June 2022.
\newblock ISSN 2469-9934.
\newblock \doi{10.1103/physreva.105.062431}.
\newblock URL \url{http://dx.doi.org/10.1103/PhysRevA.105.062431}.

\bibitem[Bu et~al.(2023)Bu, Koh, Li, Luo, and Zhang]{Bu_2023}
K.~Bu, D.~E. Koh, L.~Li, Q.~Luo, and Y.~Zhang.
\newblock Effects of quantum resources and noise on the statistical complexity of quantum circuits.
\newblock \emph{Quantum Science and Technology}, 8\penalty0 (2):\penalty0 025013, Feb. 2023.
\newblock ISSN 2058-9565.
\newblock \doi{10.1088/2058-9565/acb56a}.
\newblock URL \url{http://dx.doi.org/10.1088/2058-9565/acb56a}.

\bibitem[Bu et~al.(2024)Bu, Koh, Garcia, and Jaffe]{Bu_2024}
K.~Bu, D.~E. Koh, R.~J. Garcia, and A.~Jaffe.
\newblock Classical shadows with pauli-invariant unitary ensembles.
\newblock \emph{npj Quantum Information}, 10\penalty0 (1), Jan. 2024.
\newblock ISSN 2056-6387.
\newblock \doi{10.1038/s41534-023-00801-w}.
\newblock URL \url{http://dx.doi.org/10.1038/s41534-023-00801-w}.

\bibitem[Bultrini et~al.(2023)Bultrini, Wang, Czarnik, Gordon, Cerezo, Coles, and Cincio]{Bultrini_2023}
D.~Bultrini, S.~Wang, P.~Czarnik, M.~H. Gordon, M.~Cerezo, P.~J. Coles, and L.~Cincio.
\newblock The battle of clean and dirty qubits in the era of partial error correction.
\newblock \emph{Quantum}, 7:\penalty0 1060, July 2023.
\newblock ISSN 2521-327X.
\newblock \doi{10.22331/q-2023-07-13-1060}.
\newblock URL \url{http://dx.doi.org/10.22331/q-2023-07-13-1060}.

\bibitem[Cerezo et~al.(2021)Cerezo, Sone, Volkoff, Cincio, and Coles]{Cerezo_2021}
M.~Cerezo, A.~Sone, T.~Volkoff, L.~Cincio, and P.~J. Coles.
\newblock Cost function dependent barren plateaus in shallow parametrized quantum circuits.
\newblock \emph{Nature Communications}, 12\penalty0 (1), Mar. 2021.
\newblock ISSN 2041-1723.
\newblock \doi{10.1038/s41467-021-21728-w}.
\newblock URL \url{http://dx.doi.org/10.1038/s41467-021-21728-w}.

\bibitem[Cerezo et~al.(2022{\natexlab{a}})Cerezo, Sharma, Arrasmith, and Coles]{cerezo2022variational}
M.~Cerezo, K.~Sharma, A.~Arrasmith, and P.~J. Coles.
\newblock Variational quantum state eigensolver.
\newblock \emph{npj Quantum Information}, 8\penalty0 (1):\penalty0 113, 2022{\natexlab{a}}.

\bibitem[Cerezo et~al.(2022{\natexlab{b}})Cerezo, Verdon, Huang, Cincio, and Coles]{cerezo2022challenges}
M.~Cerezo, G.~Verdon, H.-Y. Huang, L.~Cincio, and P.~J. Coles.
\newblock Challenges and opportunities in quantum machine learning.
\newblock \emph{Nature Computational Science}, 2\penalty0 (9):\penalty0 567--576, 2022{\natexlab{b}}.

\bibitem[Cervero~Martín et~al.(2023)Cervero~Martín, Plekhanov, and Lubasch]{Cervero_Mart_n_2023}
E.~Cervero~Martín, K.~Plekhanov, and M.~Lubasch.
\newblock Barren plateaus in quantum tensor network optimization.
\newblock \emph{Quantum}, 7:\penalty0 974, Apr. 2023.
\newblock ISSN 2521-327X.
\newblock \doi{10.22331/q-2023-04-13-974}.
\newblock URL \url{http://dx.doi.org/10.22331/q-2023-04-13-974}.

\bibitem[Chang(2006)]{chang2006introduction}
H.-H. Chang.
\newblock An introduction to error-correcting codes: From classical to quantum, 2006.

\bibitem[Chen et~al.(2021)Chen, Yu, Zeng, and Flammia]{Chen_2021}
S.~Chen, W.~Yu, P.~Zeng, and S.~T. Flammia.
\newblock Robust shadow estimation.
\newblock \emph{PRX Quantum}, 2\penalty0 (3), Sept. 2021.
\newblock ISSN 2691-3399.
\newblock \doi{10.1103/prxquantum.2.030348}.
\newblock URL \url{http://dx.doi.org/10.1103/PRXQuantum.2.030348}.

\bibitem[Cheng et~al.(2018)Cheng, Chen, and Wang]{cheng2018information}
S.~Cheng, J.~Chen, and L.~Wang.
\newblock Information perspective to probabilistic modeling: Boltzmann machines versus born machines.
\newblock \emph{Entropy}, 20\penalty0 (8):\penalty0 583, 2018.

\bibitem[Chia et~al.(2020)Chia, Gilyén, Li, Lin, Tang, and Wang]{Chia_2020}
N.-H. Chia, A.~Gilyén, T.~Li, H.-H. Lin, E.~Tang, and C.~Wang.
\newblock Sampling-based sublinear low-rank matrix arithmetic framework for dequantizing quantum machine learning.
\newblock In \emph{Proceedings of the 52nd Annual ACM SIGACT Symposium on Theory of Computing}, STOC ’20. ACM, June 2020.
\newblock \doi{10.1145/3357713.3384314}.
\newblock URL \url{http://dx.doi.org/10.1145/3357713.3384314}.

\bibitem[Chizat et~al.(2020)Chizat, Oyallon, and Bach]{chizat2020lazy}
L.~Chizat, E.~Oyallon, and F.~Bach.
\newblock On lazy training in differentiable programming, 2020.

\bibitem[Ciliberto et~al.(2018)Ciliberto, Herbster, Ialongo, Pontil, Rocchetto, Severini, and Wossnig]{Ciliberto_2018}
C.~Ciliberto, M.~Herbster, A.~D. Ialongo, M.~Pontil, A.~Rocchetto, S.~Severini, and L.~Wossnig.
\newblock Quantum machine learning: a classical perspective.
\newblock \emph{Proceedings of the Royal Society A: Mathematical, Physical and Engineering Sciences}, 474\penalty0 (2209):\penalty0 20170551, Jan. 2018.
\newblock ISSN 1471-2946.
\newblock \doi{10.1098/rspa.2017.0551}.
\newblock URL \url{http://dx.doi.org/10.1098/rspa.2017.0551}.

\bibitem[Cincio et~al.(2021)Cincio, Rudinger, Sarovar, and Coles]{Cincio_2021}
L.~Cincio, K.~Rudinger, M.~Sarovar, and P.~J. Coles.
\newblock Machine learning of noise-resilient quantum circuits.
\newblock \emph{PRX Quantum}, 2\penalty0 (1), Feb. 2021.
\newblock ISSN 2691-3399.
\newblock \doi{10.1103/prxquantum.2.010324}.
\newblock URL \url{http://dx.doi.org/10.1103/PRXQuantum.2.010324}.

\bibitem[Cotler et~al.(2017)Cotler, Hunter-Jones, Liu, and Yoshida]{Cotler_2017}
J.~Cotler, N.~Hunter-Jones, J.~Liu, and B.~Yoshida.
\newblock Chaos, complexity, and random matrices.
\newblock \emph{Journal of High Energy Physics}, 2017\penalty0 (11), Nov. 2017.
\newblock ISSN 1029-8479.
\newblock \doi{10.1007/jhep11(2017)048}.
\newblock URL \url{http://dx.doi.org/10.1007/JHEP11(2017)048}.

\bibitem[Cotler et~al.(2021)Cotler, Huang, and McClean]{cotler2021revisiting}
J.~Cotler, H.-Y. Huang, and J.~R. McClean.
\newblock Revisiting dequantization and quantum advantage in learning tasks.
\newblock \emph{arXiv preprint arXiv:2112.00811}, 2021.

\bibitem[Czarnik et~al.(2021)Czarnik, Arrasmith, Coles, and Cincio]{Czarnik_2021}
P.~Czarnik, A.~Arrasmith, P.~J. Coles, and L.~Cincio.
\newblock Error mitigation with clifford quantum-circuit data.
\newblock \emph{Quantum}, 5:\penalty0 592, Nov. 2021.
\newblock ISSN 2521-327X.
\newblock \doi{10.22331/q-2021-11-26-592}.
\newblock URL \url{http://dx.doi.org/10.22331/q-2021-11-26-592}.

\bibitem[Deshpande et~al.(2022)Deshpande, Niroula, Shtanko, Gorshkov, Fefferman, and Gullans]{Deshpande_2022}
A.~Deshpande, P.~Niroula, O.~Shtanko, A.~V. Gorshkov, B.~Fefferman, and M.~J. Gullans.
\newblock Tight bounds on the convergence of noisy random circuits to the uniform distribution.
\newblock \emph{PRX Quantum}, 3\penalty0 (4), Dec. 2022.
\newblock ISSN 2691-3399.
\newblock \doi{10.1103/prxquantum.3.040329}.
\newblock URL \url{http://dx.doi.org/10.1103/PRXQuantum.3.040329}.

\bibitem[Dunjko and Wittek(2020)]{Dunjko2020Non}
V.~Dunjko and P.~Wittek.
\newblock A non-review of quantum machine learning: trends and explorations.
\newblock \emph{Quantum Views}, 4:\penalty0 32, 03 2020.
\newblock \doi{10.22331/qv-2020-03-17-32}.

\bibitem[Elben et~al.(2022)Elben, Flammia, Huang, Kueng, Preskill, Vermersch, and Zoller]{Elben_2022}
A.~Elben, S.~T. Flammia, H.-Y. Huang, R.~Kueng, J.~Preskill, B.~Vermersch, and P.~Zoller.
\newblock The randomized measurement toolbox.
\newblock \emph{Nature Reviews Physics}, 5\penalty0 (1):\penalty0 9–24, Dec. 2022.
\newblock ISSN 2522-5820.
\newblock \doi{10.1038/s42254-022-00535-2}.
\newblock URL \url{http://dx.doi.org/10.1038/s42254-022-00535-2}.

\bibitem[Endo et~al.(2021)Endo, Cai, Benjamin, and Yuan]{endo2021hybrid}
S.~Endo, Z.~Cai, S.~C. Benjamin, and X.~Yuan.
\newblock Hybrid quantum-classical algorithms and quantum error mitigation.
\newblock \emph{Journal of the Physical Society of Japan}, 90\penalty0 (3):\penalty0 032001, 2021.

\bibitem[Evans et~al.(2019)Evans, Harper, and Flammia]{evans2019scalable}
T.~J. Evans, R.~Harper, and S.~T. Flammia.
\newblock Scalable bayesian hamiltonian learning, 2019.

\bibitem[Farhi et~al.(2014)Farhi, Goldstone, and Gutmann]{farhi2014quantum}
E.~Farhi, J.~Goldstone, and S.~Gutmann.
\newblock A quantum approximate optimization algorithm.
\newblock \emph{arXiv preprint arXiv:1411.4028}, 2014.

\bibitem[Feynman et~al.(1996)Feynman, Hey, and Allen]{feynman1996lectures}
R.~Feynman, A.~Hey, and R.~Allen.
\newblock \emph{Lectures On Computation}.
\newblock Advanced book program. Basic Books, 1996.
\newblock ISBN 9780201489910.
\newblock URL \url{https://books.google.com.tw/books?id=-olQAAAAMAAJ}.

\bibitem[Garcia-Saez and Riu(2019)]{garciasaez2019quantum}
A.~Garcia-Saez and J.~Riu.
\newblock Quantum observables for continuous control of the quantum approximate optimization algorithm via reinforcement learning, 2019.

\bibitem[Gilyén et~al.(2019)Gilyén, Su, Low, and Wiebe]{Gily_n_2019}
A.~Gilyén, Y.~Su, G.~H. Low, and N.~Wiebe.
\newblock Quantum singular value transformation and beyond: exponential improvements for quantum matrix arithmetics.
\newblock In \emph{Proceedings of the 51st Annual ACM SIGACT Symposium on Theory of Computing}, STOC ’19. ACM, June 2019.
\newblock \doi{10.1145/3313276.3316366}.
\newblock URL \url{http://dx.doi.org/10.1145/3313276.3316366}.

\bibitem[Granade et~al.(2012)Granade, Ferrie, Wiebe, and Cory]{Granade_2012}
C.~E. Granade, C.~Ferrie, N.~Wiebe, and D.~G. Cory.
\newblock Robust online hamiltonian learning.
\newblock \emph{New Journal of Physics}, 14\penalty0 (10):\penalty0 103013, Oct. 2012.
\newblock ISSN 1367-2630.
\newblock \doi{10.1088/1367-2630/14/10/103013}.
\newblock URL \url{http://dx.doi.org/10.1088/1367-2630/14/10/103013}.

\bibitem[Grimsley et~al.(2019)Grimsley, Economou, Barnes, and Mayhall]{grimsley2019adaptive}
H.~R. Grimsley, S.~E. Economou, E.~Barnes, and N.~J. Mayhall.
\newblock An adaptive variational algorithm for exact molecular simulations on a quantum computer.
\newblock \emph{Nature communications}, 10\penalty0 (1):\penalty0 3007, 2019.

\bibitem[Grover(1996)]{ChairmanMiller1996Proceedings}
L.~K. Grover.
\newblock A fast quantum mechanical algorithm for database search, 1996.

\bibitem[Gustafsson et~al.(2014)Gustafsson, Aref, Kockum, Ekström, Johansson, and Delsing]{Gustafsson_2014}
M.~V. Gustafsson, T.~Aref, A.~F. Kockum, M.~K. Ekström, G.~Johansson, and P.~Delsing.
\newblock Propagating phonons coupled to an artificial atom.
\newblock \emph{Science}, 346\penalty0 (6206):\penalty0 207–211, 2014.
\newblock ISSN 1095-9203.
\newblock \doi{10.1126/science.1257219}.
\newblock URL \url{http://dx.doi.org/10.1126/science.1257219}.

\bibitem[HAFFNER et~al.(2008)HAFFNER, ROOS, and BLATT]{HAFFNER_2008}
H.~HAFFNER, C.~ROOS, and R.~BLATT.
\newblock Quantum computing with trapped ions.
\newblock \emph{Physics Reports}, 469\penalty0 (4):\penalty0 155--203, dec 2008.
\newblock \doi{10.1016/j.physrep.2008.09.003}.
\newblock URL \url{https://doi.org/10.1016%2Fj.physrep.2008.09.003}.

\bibitem[Hakkaku et~al.(2022)Hakkaku, Tashima, Mitarai, Mizukami, and Fujii]{Hakkaku_2022}
S.~Hakkaku, Y.~Tashima, K.~Mitarai, W.~Mizukami, and K.~Fujii.
\newblock Quantifying fermionic nonlinearity of quantum circuits.
\newblock \emph{Physical Review Research}, 4\penalty0 (4), Nov. 2022.
\newblock ISSN 2643-1564.
\newblock \doi{10.1103/physrevresearch.4.043100}.
\newblock URL \url{http://dx.doi.org/10.1103/PhysRevResearch.4.043100}.

\bibitem[Hann(2021)]{hann2021practicality}
C.~T. Hann.
\newblock \emph{Practicality of Quantum Random Access Memory}.
\newblock PhD thesis, Yale University, 2021.

\bibitem[Hann et~al.(2019{\natexlab{a}})Hann, Zou, Zhang, Chu, Schoelkopf, Girvin, and Jiang]{Hann_2019}
C.~T. Hann, C.-L. Zou, Y.~Zhang, Y.~Chu, R.~J. Schoelkopf, S.~Girvin, and L.~Jiang.
\newblock Hardware-efficient quantum random access memory with hybrid quantum acoustic systems.
\newblock \emph{Physical Review Letters}, 123\penalty0 (25), 2019{\natexlab{a}}.
\newblock ISSN 1079-7114.
\newblock \doi{10.1103/physrevlett.123.250501}.
\newblock URL \url{http://dx.doi.org/10.1103/PhysRevLett.123.250501}.

\bibitem[Hann et~al.(2019{\natexlab{b}})Hann, Zou, Zhang, Chu, Schoelkopf, Girvin, and Jiang]{Hann1}
C.~T. Hann, C.-L. Zou, Y.~Zhang, Y.~Chu, R.~J. Schoelkopf, S.~M. Girvin, and L.~Jiang.
\newblock Hardware-efficient quantum random access memory with hybrid quantum acoustic systems.
\newblock \emph{Phys. Rev. Lett.}, 123:\penalty0 250501, Dec 2019{\natexlab{b}}.

\bibitem[Harrow and Napp(2021)]{Harrow_2021Low}
A.~W. Harrow and J.~C. Napp.
\newblock Low-depth gradient measurements can improve convergence in variational hybrid quantum-classical algorithms.
\newblock \emph{Physical Review Letters}, 126\penalty0 (14), apr 2021.
\newblock \doi{10.1103/physrevlett.126.140502}.
\newblock URL \url{https://doi.org/10.1103%2Fphysrevlett.126.140502}.

\bibitem[Harrow et~al.(2009)Harrow, Hassidim, and Lloyd]{harrow2009quantum}
A.~W. Harrow, A.~Hassidim, and S.~Lloyd.
\newblock Quantum algorithm for linear systems of equations.
\newblock \emph{Physical review letters}, 103\penalty0 (15):\penalty0 150502, 2009.

\bibitem[Harrow et~al.(2017)Harrow, Lin, and Montanaro]{Harrow_2017}
A.~W. Harrow, C.~Y.-Y. Lin, and A.~Montanaro.
\newblock Sequential measurements, disturbance and property testing.
\newblock In \emph{Proceedings of the Twenty-Eighth Annual ACM-SIAM Symposium on Discrete Algorithms}. Society for Industrial and Applied Mathematics, Jan. 2017.
\newblock \doi{10.1137/1.9781611974782.105}.
\newblock URL \url{http://dx.doi.org/10.1137/1.9781611974782.105}.

\bibitem[Holmes et~al.(2022)Holmes, Sharma, Cerezo, and Coles]{Holmes_2022}
Z.~Holmes, K.~Sharma, M.~Cerezo, and P.~J. Coles.
\newblock Connecting ansatz expressibility to gradient magnitudes and barren plateaus.
\newblock \emph{PRX Quantum}, 3\penalty0 (1), Jan. 2022.
\newblock ISSN 2691-3399.
\newblock \doi{10.1103/prxquantum.3.010313}.
\newblock URL \url{http://dx.doi.org/10.1103/PRXQuantum.3.010313}.

\bibitem[{Huang} et~al.(2019){Huang}, {Bharti}, and {Rebentrost}]{Huang2019Near}
H.-Y. {Huang}, K.~{Bharti}, and P.~{Rebentrost}.
\newblock {Near-term quantum algorithms for linear systems of equations}.
\newblock \emph{arXiv e-prints}, art. arXiv:1909.07344, Sept. 2019.
\newblock \doi{10.48550/arXiv.1909.07344}.

\bibitem[Huang et~al.(2020{\natexlab{a}})Huang, Kueng, and Preskill]{Huang2020Predeicting}
H.-Y. Huang, R.~Kueng, and J.~Preskill.
\newblock Predicting many properties of a quantum system from very few measurements.
\newblock \emph{Nature Physics}, 16\penalty0 (10):\penalty0 1050--1057, jun 2020{\natexlab{a}}.
\newblock \doi{10.1038/s41567-020-0932-7}.
\newblock URL \url{https://doi.org/10.1038%2Fs41567-020-0932-7}.

\bibitem[Huang et~al.(2020{\natexlab{b}})Huang, Kueng, and Preskill]{Huang_2020}
H.-Y. Huang, R.~Kueng, and J.~Preskill.
\newblock Predicting many properties of a quantum system from very few measurements.
\newblock \emph{Nature Physics}, 16\penalty0 (10):\penalty0 1050–1057, June 2020{\natexlab{b}}.
\newblock ISSN 1745-2481.
\newblock \doi{10.1038/s41567-020-0932-7}.
\newblock URL \url{http://dx.doi.org/10.1038/s41567-020-0932-7}.

\bibitem[Huang et~al.(2021{\natexlab{a}})Huang, Broughton, Mohseni, Babbush, Boixo, Neven, and McClean]{Huang_2021}
H.-Y. Huang, M.~Broughton, M.~Mohseni, R.~Babbush, S.~Boixo, H.~Neven, and J.~R. McClean.
\newblock Power of data in quantum machine learning.
\newblock \emph{Nature Communications}, 12\penalty0 (1), may 2021{\natexlab{a}}.
\newblock \doi{10.1038/s41467-021-22539-9}.
\newblock URL \url{https://doi.org/10.1038%2Fs41467-021-22539-9}.

\bibitem[Huang et~al.(2021{\natexlab{b}})Huang, Kueng, and Preskill]{Huang2021Efficient}
H.-Y. Huang, R.~Kueng, and J.~Preskill.
\newblock Efficient estimation of pauli observables by derandomization.
\newblock \emph{Phys. Rev. Lett.}, 127:\penalty0 030503, Jul 2021{\natexlab{b}}.
\newblock \doi{10.1103/PhysRevLett.127.030503}.
\newblock URL \url{https://link.aps.org/doi/10.1103/PhysRevLett.127.030503}.

\bibitem[Huang et~al.(2021{\natexlab{c}})Huang, Kueng, and Preskill]{huang2021information}
H.-Y. Huang, R.~Kueng, and J.~Preskill.
\newblock Information-theoretic bounds on quantum advantage in machine learning.
\newblock \emph{Physical Review Letters}, 126\penalty0 (19):\penalty0 190505, 2021{\natexlab{c}}.

\bibitem[Huang et~al.(2022{\natexlab{a}})Huang, Broughton, Cotler, Chen, Li, Mohseni, Neven, Babbush, Kueng, Preskill, et~al.]{huang2022quantum}
H.-Y. Huang, M.~Broughton, J.~Cotler, S.~Chen, J.~Li, M.~Mohseni, H.~Neven, R.~Babbush, R.~Kueng, J.~Preskill, et~al.
\newblock Quantum advantage in learning from experiments.
\newblock \emph{Science}, 376\penalty0 (6598):\penalty0 1182--1186, 2022{\natexlab{a}}.

\bibitem[Huang et~al.(2022{\natexlab{b}})Huang, Kueng, Torlai, Albert, and Preskill]{Huang_2022}
H.-Y. Huang, R.~Kueng, G.~Torlai, V.~V. Albert, and J.~Preskill.
\newblock Provably efficient machine learning for quantum many-body problems.
\newblock \emph{Science}, 377\penalty0 (6613), Sept. 2022{\natexlab{b}}.
\newblock ISSN 1095-9203.
\newblock \doi{10.1126/science.abk3333}.
\newblock URL \url{http://dx.doi.org/10.1126/science.abk3333}.

\bibitem[Jain and Kar(2017)]{Jain_2017}
P.~Jain and P.~Kar.
\newblock Non-convex optimization for machine learning.
\newblock \emph{Foundations and Trends® in Machine Learning}, 10\penalty0 (3–4):\penalty0 142–336, 2017.
\newblock ISSN 1935-8245.
\newblock \doi{10.1561/2200000058}.
\newblock URL \url{http://dx.doi.org/10.1561/2200000058}.

\bibitem[Jin et~al.(2019)Jin, Netrapalli, Ge, Kakade, and Jordan]{jin2019nonconvex}
C.~Jin, P.~Netrapalli, R.~Ge, S.~M. Kakade, and M.~I. Jordan.
\newblock On nonconvex optimization for machine learning: Gradients, stochasticity, and saddle points, 2019.

\bibitem[Kandala et~al.(2017)Kandala, Mezzacapo, Temme, Takita, Brink, Chow, and Gambetta]{kandala2017hardware}
A.~Kandala, A.~Mezzacapo, K.~Temme, M.~Takita, M.~Brink, J.~M. Chow, and J.~M. Gambetta.
\newblock Hardware-efficient variational quantum eigensolver for small molecules and quantum magnets.
\newblock \emph{nature}, 549\penalty0 (7671):\penalty0 242--246, 2017.

\bibitem[Kerenidis and Prakash(2016)]{kerenidis2016quantum}
I.~Kerenidis and A.~Prakash.
\newblock Quantum recommendation systems, 2016.

\bibitem[Kiani et~al.(2020)Kiani, Lloyd, and Maity]{kiani2020learning}
B.~T. Kiani, S.~Lloyd, and R.~Maity.
\newblock Learning unitaries by gradient descent, 2020.

\bibitem[Kivlichan et~al.(2018)Kivlichan, McClean, Wiebe, Gidney, Aspuru-Guzik, Chan, and Babbush]{kivlichan2018quantum}
I.~D. Kivlichan, J.~McClean, N.~Wiebe, C.~Gidney, A.~Aspuru-Guzik, G.~K.-L. Chan, and R.~Babbush.
\newblock Quantum simulation of electronic structure with linear depth and connectivity.
\newblock \emph{Physical review letters}, 120\penalty0 (11):\penalty0 110501, 2018.

\bibitem[Koh and Grewal(2022)]{Koh_2022}
D.~E. Koh and S.~Grewal.
\newblock Classical shadows with noise.
\newblock \emph{Quantum}, 6:\penalty0 776, Aug. 2022.
\newblock ISSN 2521-327X.
\newblock \doi{10.22331/q-2022-08-16-776}.
\newblock URL \url{http://dx.doi.org/10.22331/q-2022-08-16-776}.

\bibitem[Kottmann et~al.(2021)Kottmann, Krenn, Kyaw, Alperin-Lea, and Aspuru-Guzik]{kottmann2021quantum}
J.~S. Kottmann, M.~Krenn, T.~H. Kyaw, S.~Alperin-Lea, and A.~Aspuru-Guzik.
\newblock Quantum computer-aided design of quantum optics hardware.
\newblock \emph{Quantum Science and Technology}, 6\penalty0 (3):\penalty0 035010, 2021.

\bibitem[Krantz et~al.(2019)Krantz, Kjaergaard, Yan, Orlando, Gustavsson, and Oliver]{krantz2019quantum}
P.~Krantz, M.~Kjaergaard, F.~Yan, T.~P. Orlando, S.~Gustavsson, and W.~D. Oliver.
\newblock A quantum engineer's guide to superconducting qubits.
\newblock \emph{Applied physics reviews}, 6\penalty0 (2), 2019.

\bibitem[Krenn et~al.(2020)Krenn, Erhard, and Zeilinger]{krenn2020computer}
M.~Krenn, M.~Erhard, and A.~Zeilinger.
\newblock Computer-inspired quantum experiments.
\newblock \emph{Nature Reviews Physics}, 2\penalty0 (11):\penalty0 649--661, 2020.

\bibitem[Larocca et~al.(2023)Larocca, Ju, Garcia-Martin, Coles, and Cerezo]{Larocca_2023}
M.~Larocca, N.~Ju, D.~Garcia-Martin, P.~J. Coles, and M.~Cerezo.
\newblock Theory of overparametrization in quantum neural networks.
\newblock \emph{Nature Computational Science}, 3\penalty0 (6):\penalty0 542–551, 2023.
\newblock ISSN 2662-8457.
\newblock \doi{10.1038/s43588-023-00467-6}.
\newblock URL \url{http://dx.doi.org/10.1038/s43588-023-00467-6}.

\bibitem[LaRose and Coyle(2020)]{LaRose_2020}
R.~LaRose and B.~Coyle.
\newblock Robust data encodings for quantum classifiers.
\newblock \emph{Physical Review A}, 102\penalty0 (3), Sept. 2020.
\newblock ISSN 2469-9934.
\newblock \doi{10.1103/physreva.102.032420}.
\newblock URL \url{http://dx.doi.org/10.1103/PhysRevA.102.032420}.

\bibitem[LaRose et~al.(2019)LaRose, Tikku, O’Neel-Judy, Cincio, and Coles]{LaRose_2019}
R.~LaRose, A.~Tikku, E.~O’Neel-Judy, L.~Cincio, and P.~J. Coles.
\newblock Variational quantum state diagonalization.
\newblock \emph{npj Quantum Information}, 5\penalty0 (1), 2019.
\newblock ISSN 2056-6387.
\newblock \doi{10.1038/s41534-019-0167-6}.
\newblock URL \url{http://dx.doi.org/10.1038/s41534-019-0167-6}.

\bibitem[Lee et~al.(2018)Lee, Huggins, Head-Gordon, and Whaley]{lee2018generalized}
J.~Lee, W.~J. Huggins, M.~Head-Gordon, and K.~B. Whaley.
\newblock Generalized unitary coupled cluster wave functions for quantum computation.
\newblock \emph{Journal of chemical theory and computation}, 15\penalty0 (1):\penalty0 311--324, 2018.

\bibitem[Li et~al.(2020)Li, Fan, Coram, Riley, Leichenauer, et~al.]{li2020quantum}
L.~Li, M.~Fan, M.~Coram, P.~Riley, S.~Leichenauer, et~al.
\newblock Quantum optimization with a novel gibbs objective function and ansatz architecture search.
\newblock \emph{Physical Review Research}, 2\penalty0 (2):\penalty0 023074, 2020.

\bibitem[Liu(2018)]{Liu_2018}
J.~Liu.
\newblock Spectral form factors and late time quantum chaos.
\newblock \emph{Physical Review D}, 98\penalty0 (8), Oct. 2018.
\newblock ISSN 2470-0029.
\newblock \doi{10.1103/physrevd.98.086026}.
\newblock URL \url{http://dx.doi.org/10.1103/PhysRevD.98.086026}.

\bibitem[Liu(2020)]{Liu_2020}
J.~Liu.
\newblock Scrambling and decoding the charged quantum information.
\newblock \emph{Physical Review Research}, 2\penalty0 (4), Oct. 2020.
\newblock ISSN 2643-1564.
\newblock \doi{10.1103/physrevresearch.2.043164}.
\newblock URL \url{http://dx.doi.org/10.1103/PhysRevResearch.2.043164}.

\bibitem[Liu et~al.(2022{\natexlab{a}})Liu, Lin, and Jiang]{liu2022laziness}
J.~Liu, Z.~Lin, and L.~Jiang.
\newblock Laziness, barren plateau, and noise in machine learning, 2022{\natexlab{a}}.

\bibitem[Liu et~al.(2022{\natexlab{b}})Liu, Tacchino, Glick, Jiang, and Mezzacapo]{liu2022representation}
J.~Liu, F.~Tacchino, J.~R. Glick, L.~Jiang, and A.~Mezzacapo.
\newblock Representation learning via quantum neural tangent kernels.
\newblock \emph{PRX Quantum}, 3\penalty0 (3):\penalty0 030323, 2022{\natexlab{b}}.

\bibitem[Liu et~al.(2022{\natexlab{c}})Liu, Wilde, Mele, Jiang, and Eisert]{liu2022noise}
J.~Liu, F.~Wilde, A.~A. Mele, L.~Jiang, and J.~Eisert.
\newblock Noise can be helpful for variational quantum algorithms.
\newblock \emph{arXiv preprint arXiv:2210.06723}, 2022{\natexlab{c}}.

\bibitem[Liu et~al.(2023)Liu, Liu, Liu, Ye, Wang, Alexeev, Eisert, and Jiang]{liu2023towards}
J.~Liu, M.~Liu, J.-P. Liu, Z.~Ye, Y.~Wang, Y.~Alexeev, J.~Eisert, and L.~Jiang.
\newblock Towards provably efficient quantum algorithms for large-scale machine-learning models.
\newblock \emph{arXiv preprint arXiv:2303.03428}, 2023.

\bibitem[Liu et~al.(2021)Liu, Kolden, Krovi, Loureiro, Trivisa, and Childs]{Liu2021Efficient}
J.-P. Liu, H.~{\O}. Kolden, H.~K. Krovi, N.~F. Loureiro, K.~Trivisa, and A.~M. Childs.
\newblock Efficient quantum algorithm for dissipative nonlinear differential equations.
\newblock \emph{Proceedings of the National Academy of Sciences}, 118\penalty0 (35), 2021.
\newblock ISSN 1091-6490.
\newblock \doi{10.1073/pnas.2026805118}.
\newblock URL \url{http://dx.doi.org/10.1073/pnas.2026805118}.

\bibitem[Lloyd(1996)]{lloyd1996universal}
S.~Lloyd.
\newblock Universal quantum simulators.
\newblock \emph{Science}, 273\penalty0 (5278):\penalty0 1073--1078, 1996.

\bibitem[Lloyd et~al.(2013)Lloyd, Mohseni, and Rebentrost]{lloyd2013quantum}
S.~Lloyd, M.~Mohseni, and P.~Rebentrost.
\newblock Quantum algorithms for supervised and unsupervised machine learning, 2013.

\bibitem[Lloyd et~al.(2014)Lloyd, Mohseni, and Rebentrost]{Lloyd_2014}
S.~Lloyd, M.~Mohseni, and P.~Rebentrost.
\newblock Quantum principal component analysis.
\newblock \emph{Nature Physics}, 10\penalty0 (9):\penalty0 631–633, July 2014.
\newblock ISSN 1745-2481.
\newblock \doi{10.1038/nphys3029}.
\newblock URL \url{http://dx.doi.org/10.1038/nphys3029}.

\bibitem[Lubasch et~al.(2020)Lubasch, Joo, Moinier, Kiffner, and Jaksch]{Lubasch_2020}
M.~Lubasch, J.~Joo, P.~Moinier, M.~Kiffner, and D.~Jaksch.
\newblock Variational quantum algorithms for nonlinear problems.
\newblock \emph{Physical Review A}, 101\penalty0 (1), Jan. 2020.
\newblock ISSN 2469-9934.
\newblock \doi{10.1103/physreva.101.010301}.
\newblock URL \url{http://dx.doi.org/10.1103/PhysRevA.101.010301}.

\bibitem[Marrero et~al.(2021)Marrero, Kieferová, and Wiebe]{marrero2021entanglement}
C.~O. Marrero, M.~Kieferová, and N.~Wiebe.
\newblock Entanglement induced barren plateaus, 2021.

\bibitem[Marvian and Lloyd(2016)]{marvian2016universal}
I.~Marvian and S.~Lloyd.
\newblock Universal quantum emulator, 2016.

\bibitem[McClean et~al.(2016)McClean, Romero, Babbush, and Aspuru-Guzik]{mcclean2016theory}
J.~R. McClean, J.~Romero, R.~Babbush, and A.~Aspuru-Guzik.
\newblock The theory of variational hybrid quantum-classical algorithms.
\newblock \emph{New Journal of Physics}, 18\penalty0 (2):\penalty0 023023, 2016.

\bibitem[McClean et~al.(2018)McClean, Boixo, Smelyanskiy, Babbush, and Neven]{McClean_2018}
J.~R. McClean, S.~Boixo, V.~N. Smelyanskiy, R.~Babbush, and H.~Neven.
\newblock Barren plateaus in quantum neural network training landscapes.
\newblock \emph{Nature Communications}, 9\penalty0 (1), Nov. 2018.
\newblock ISSN 2041-1723.
\newblock \doi{10.1038/s41467-018-07090-4}.
\newblock URL \url{http://dx.doi.org/10.1038/s41467-018-07090-4}.

\bibitem[Mitarai and Fujii(2019)]{Mitarai2019Methodology}
K.~Mitarai and K.~Fujii.
\newblock Methodology for replacing indirect measurements with direct measurements.
\newblock \emph{Physical Review Research}, 1\penalty0 (1), aug 2019.
\newblock \doi{10.1103/physrevresearch.1.013006}.
\newblock URL \url{https://doi.org/10.1103%2Fphysrevresearch.1.013006}.

\bibitem[Moll et~al.(2018)Moll, Barkoutsos, Bishop, Chow, Cross, Egger, Filipp, Fuhrer, Gambetta, Ganzhorn, et~al.]{moll2018quantum}
N.~Moll, P.~Barkoutsos, L.~S. Bishop, J.~M. Chow, A.~Cross, D.~J. Egger, S.~Filipp, A.~Fuhrer, J.~M. Gambetta, M.~Ganzhorn, et~al.
\newblock Quantum optimization using variational algorithms on near-term quantum devices.
\newblock \emph{Quantum Science and Technology}, 3\penalty0 (3):\penalty0 030503, 2018.

\bibitem[Nakanishi et~al.(2020)Nakanishi, Fujii, and Todo]{Nakanishi_2020}
K.~M. Nakanishi, K.~Fujii, and S.~Todo.
\newblock Sequential minimal optimization for quantum-classical hybrid algorithms.
\newblock \emph{Physical Review Research}, 2\penalty0 (4), oct 2020.
\newblock \doi{10.1103/physrevresearch.2.043158}.
\newblock URL \url{https://doi.org/10.1103%2Fphysrevresearch.2.043158}.

\bibitem[Nielsen and Chuang(2010)]{nielsen_chuang_2010}
M.~A. Nielsen and I.~L. Chuang.
\newblock \emph{Quantum Computation and Quantum Information: 10th Anniversary Edition}.
\newblock Cambridge University Press, 2010.
\newblock \doi{10.1017/CBO9780511976667}.

\bibitem[O'Connell et~al.(2010)O'Connell, Hofheinz, Ansmann, Bialczak, Lenander, Lucero, Neeley, Sank, Wang, Weides, Wenner, Martinis, and Cleland]{Connell2010quantum}
A.~O'Connell, M.~Hofheinz, M.~Ansmann, R.~C. Bialczak, M.~Lenander, E.~Lucero, M.~Neeley, D.~Sank, H.~Wang, M.~Weides, J.~Wenner, J.~M. Martinis, and A.~Cleland.
\newblock Quantum ground state and single-phonon control of a mechanical resonator.
\newblock \emph{Nature}, 464\penalty0 (7289):\penalty0 697—703, April 2010.
\newblock ISSN 0028-0836.
\newblock \doi{10.1038/nature08967}.
\newblock URL \url{https://doi.org/10.1038/nature08967}.

\bibitem[O'Donnell and Wright(2015)]{odonnell2015efficient}
R.~O'Donnell and J.~Wright.
\newblock Efficient quantum tomography, 2015.

\bibitem[Ostaszewski et~al.(2021)Ostaszewski, Grant, and Benedetti]{Ostaszewski_2021}
M.~Ostaszewski, E.~Grant, and M.~Benedetti.
\newblock Structure optimization for parameterized quantum circuits.
\newblock \emph{Quantum}, 5:\penalty0 391, jan 2021.
\newblock \doi{10.22331/q-2021-01-28-391}.
\newblock URL \url{https://doi.org/10.22331%2Fq-2021-01-28-391}.

\bibitem[Patti et~al.(2021)Patti, Najafi, Gao, and Yelin]{Patti_2021}
T.~L. Patti, K.~Najafi, X.~Gao, and S.~F. Yelin.
\newblock Entanglement devised barren plateau mitigation.
\newblock \emph{Physical Review Research}, 3\penalty0 (3), July 2021.
\newblock ISSN 2643-1564.
\newblock \doi{10.1103/physrevresearch.3.033090}.
\newblock URL \url{http://dx.doi.org/10.1103/PhysRevResearch.3.033090}.

\bibitem[P{\'e}rez-Salinas et~al.(2020)P{\'e}rez-Salinas, Cervera-Lierta, Gil-Fuster, and Latorre]{perez2020data}
A.~P{\'e}rez-Salinas, A.~Cervera-Lierta, E.~Gil-Fuster, and J.~I. Latorre.
\newblock Data re-uploading for a universal quantum classifier.
\newblock \emph{Quantum}, 4:\penalty0 226, 2020.

\bibitem[Peruzzo et~al.(2014)Peruzzo, McClean, Shadbolt, Yung, Zhou, Love, Aspuru-Guzik, and O’brien]{peruzzo2014variational}
A.~Peruzzo, J.~McClean, P.~Shadbolt, M.-H. Yung, X.-Q. Zhou, P.~J. Love, A.~Aspuru-Guzik, and J.~L. O’brien.
\newblock A variational eigenvalue solver on a photonic quantum processor.
\newblock \emph{Nature communications}, 5\penalty0 (1):\penalty0 4213, 2014.

\bibitem[Platt(1998)]{platt1998sequential}
J.~Platt.
\newblock Sequential minimal optimization: A fast algorithm for training support vector machines.
\newblock Technical Report MSR-TR-98-14, Microsoft, April 1998.

\bibitem[Preskill(2018)]{preskill2018quantum}
J.~Preskill.
\newblock Quantum computing in the nisq era and beyond.
\newblock \emph{Quantum}, 2:\penalty0 79, 2018.

\bibitem[Roberts and Yoshida(2017)]{Roberts_2017}
D.~A. Roberts and B.~Yoshida.
\newblock Chaos and complexity by design.
\newblock \emph{Journal of High Energy Physics}, 2017\penalty0 (4), Apr. 2017.
\newblock ISSN 1029-8479.
\newblock \doi{10.1007/jhep04(2017)121}.
\newblock URL \url{http://dx.doi.org/10.1007/JHEP04(2017)121}.

\bibitem[Roberts et~al.(2022)Roberts, Yaida, and Hanin]{roberts2022principles}
D.~A. Roberts, S.~Yaida, and B.~Hanin.
\newblock \emph{The principles of deep learning theory}.
\newblock Cambridge University Press Cambridge, MA, USA, 2022.

\bibitem[Roffe(2019)]{Roffe_2019}
J.~Roffe.
\newblock Quantum error correction: an introductory guide.
\newblock \emph{Contemporary Physics}, 60\penalty0 (3):\penalty0 226–245, July 2019.
\newblock ISSN 1366-5812.
\newblock \doi{10.1080/00107514.2019.1667078}.
\newblock URL \url{http://dx.doi.org/10.1080/00107514.2019.1667078}.

\bibitem[Sakurai and Napolitano(2020)]{Sakurai:2011zz}
J.~J. Sakurai and J.~Napolitano.
\newblock \emph{{Modern Quantum Mechanics}}.
\newblock Quantum physics, quantum information and quantum computation. Cambridge University Press, 10 2020.
\newblock ISBN 978-0-8053-8291-4, 978-1-108-52742-2, 978-1-108-58728-0.
\newblock \doi{10.1017/9781108587280}.

\bibitem[Sharma et~al.(2020)Sharma, Khatri, Cerezo, and Coles]{Sharma_2020}
K.~Sharma, S.~Khatri, M.~Cerezo, and P.~J. Coles.
\newblock Noise resilience of variational quantum compiling.
\newblock \emph{New Journal of Physics}, 22\penalty0 (4):\penalty0 043006, Apr. 2020.
\newblock ISSN 1367-2630.
\newblock \doi{10.1088/1367-2630/ab784c}.
\newblock URL \url{http://dx.doi.org/10.1088/1367-2630/ab784c}.

\bibitem[Sharma et~al.(2022)Sharma, Cerezo, Cincio, and Coles]{Sharma_2022}
K.~Sharma, M.~Cerezo, L.~Cincio, and P.~J. Coles.
\newblock Trainability of dissipative perceptron-based quantum neural networks.
\newblock \emph{Physical Review Letters}, 128\penalty0 (18), May 2022.
\newblock ISSN 1079-7114.
\newblock \doi{10.1103/physrevlett.128.180505}.
\newblock URL \url{http://dx.doi.org/10.1103/PhysRevLett.128.180505}.

\bibitem[Shor(1994)]{Shor1994Algorithm}
P.~Shor.
\newblock Algorithms for quantum computation: discrete logarithms and factoring.
\newblock In \emph{Proceedings 35th Annual Symposium on Foundations of Computer Science}, pages 124--134, 1994.
\newblock \doi{10.1109/SFCS.1994.365700}.

\bibitem[Song et~al.(2017)Song, Xu, Liu, Yang, Zheng, Deng, Xie, Huang, Guo, Zhang, Zhang, Xu, Zheng, Zhu, Wang, Chen, Lu, Han, and Pan]{Song_2017}
C.~Song, K.~Xu, W.~Liu, C.-p. Yang, S.-B. Zheng, H.~Deng, Q.~Xie, K.~Huang, Q.~Guo, L.~Zhang, P.~Zhang, D.~Xu, D.~Zheng, X.~Zhu, H.~Wang, Y.-A. Chen, C.-Y. Lu, S.~Han, and J.-W. Pan.
\newblock 10-qubit entanglement and parallel logic operations with a superconducting circuit.
\newblock \emph{Physical Review Letters}, 119\penalty0 (18), Nov. 2017.
\newblock ISSN 1079-7114.
\newblock \doi{10.1103/physrevlett.119.180511}.
\newblock URL \url{http://dx.doi.org/10.1103/PhysRevLett.119.180511}.

\bibitem[Struchalin et~al.(2021)Struchalin, Zagorovskii, Kovlakov, Straupe, and Kulik]{Struchalin2021Experimental}
G.~Struchalin, Y.~A. Zagorovskii, E.~Kovlakov, S.~Straupe, and S.~Kulik.
\newblock Experimental estimation of quantum state properties from classical shadows.
\newblock \emph{PRX Quantum}, 2:\penalty0 010307, Jan 2021.
\newblock \doi{10.1103/PRXQuantum.2.010307}.
\newblock URL \url{https://link.aps.org/doi/10.1103/PRXQuantum.2.010307}.

\bibitem[{Suzuki}(1976)]{1976CMaPh..51..183S}
M.~{Suzuki}.
\newblock {Generalized Trotter's formula and systematic approximants of exponential operators and inner derivations with applications to many-body problems}.
\newblock \emph{Communications in Mathematical Physics}, 51\penalty0 (2):\penalty0 183--190, June 1976.
\newblock \doi{10.1007/BF01609348}.

\bibitem[Tang(2019)]{tang2019quantum}
E.~Tang.
\newblock A quantum-inspired classical algorithm for recommendation systems.
\newblock In \emph{Proceedings of the 51st annual ACM SIGACT symposium on theory of computing}, pages 217--228, 2019.

\bibitem[Tang(2021)]{Tang_2021}
E.~Tang.
\newblock Quantum principal component analysis only achieves an exponential speedup because of its state preparation assumptions.
\newblock \emph{Physical Review Letters}, 127\penalty0 (6), Aug. 2021.
\newblock ISSN 1079-7114.
\newblock \doi{10.1103/physrevlett.127.060503}.
\newblock URL \url{http://dx.doi.org/10.1103/PhysRevLett.127.060503}.

\bibitem[{Taube} and {Bartlett}(2006)]{Taube2006New}
A.~G. {Taube} and R.~J. {Bartlett}.
\newblock {New perspectives on unitary coupled-cluster theory}.
\newblock \emph{International Journal of Quantum Chemistry}, 106\penalty0 (15):\penalty0 3393--3401, Jan. 2006.
\newblock \doi{10.1002/qua.21198}.

\bibitem[Temme et~al.(2017)Temme, Bravyi, and Gambetta]{Temme_2017}
K.~Temme, S.~Bravyi, and J.~M. Gambetta.
\newblock Error mitigation for short-depth quantum circuits.
\newblock \emph{Physical Review Letters}, 119\penalty0 (18), Nov. 2017.
\newblock ISSN 1079-7114.
\newblock \doi{10.1103/physrevlett.119.180509}.
\newblock URL \url{http://dx.doi.org/10.1103/PhysRevLett.119.180509}.

\bibitem[Uvarov and Biamonte(2021)]{Uvarov_2021}
A.~V. Uvarov and J.~D. Biamonte.
\newblock On barren plateaus and cost function locality in variational quantum algorithms.
\newblock \emph{Journal of Physics A: Mathematical and Theoretical}, 54\penalty0 (24):\penalty0 245301, May 2021.
\newblock ISSN 1751-8121.
\newblock \doi{10.1088/1751-8121/abfac7}.
\newblock URL \url{http://dx.doi.org/10.1088/1751-8121/abfac7}.

\bibitem[Verdon et~al.(2019)Verdon, Broughton, McClean, Sung, Babbush, Jiang, Neven, and Mohseni]{verdon2019learning}
G.~Verdon, M.~Broughton, J.~R. McClean, K.~J. Sung, R.~Babbush, Z.~Jiang, H.~Neven, and M.~Mohseni.
\newblock Learning to learn with quantum neural networks via classical neural networks, 2019.

\bibitem[Wang et~al.(2021{\natexlab{a}})Wang, Czarnik, Arrasmith, Cerezo, Cincio, and Coles]{wang2021error}
S.~Wang, P.~Czarnik, A.~Arrasmith, M.~Cerezo, L.~Cincio, and P.~J. Coles.
\newblock Can error mitigation improve trainability of noisy variational quantum algorithms?, 2021{\natexlab{a}}.

\bibitem[Wang et~al.(2021{\natexlab{b}})Wang, Fontana, Cerezo, Sharma, Sone, Cincio, and Coles]{Wang_2021}
S.~Wang, E.~Fontana, M.~Cerezo, K.~Sharma, A.~Sone, L.~Cincio, and P.~J. Coles.
\newblock Noise-induced barren plateaus in variational quantum algorithms.
\newblock \emph{Nature Communications}, 12\penalty0 (1), Nov. 2021{\natexlab{b}}.
\newblock ISSN 2041-1723.
\newblock \doi{10.1038/s41467-021-27045-6}.
\newblock URL \url{http://dx.doi.org/10.1038/s41467-021-27045-6}.

\bibitem[Wang et~al.(2023)Wang, Alexeev, Jiang, Chong, and Liu]{wang2023fundamental}
Y.~Wang, Y.~Alexeev, L.~Jiang, F.~T. Chong, and J.~Liu.
\newblock Fundamental causal bounds of quantum random access memories.
\newblock \emph{arXiv preprint arXiv:2307.13460}, 2023.

\bibitem[Wauters et~al.(2020)Wauters, Panizon, Mbeng, and Santoro]{Wauters_2020Reinforcement}
M.~M. Wauters, E.~Panizon, G.~B. Mbeng, and G.~E. Santoro.
\newblock Reinforcement-learning-assisted quantum optimization.
\newblock \emph{Physical Review Research}, 2\penalty0 (3), sep 2020.
\newblock \doi{10.1103/physrevresearch.2.033446}.
\newblock URL \url{https://doi.org/10.1103%2Fphysrevresearch.2.033446}.

\bibitem[Wecker et~al.(2015)Wecker, Hastings, and Troyer]{wecker2015progress}
D.~Wecker, M.~B. Hastings, and M.~Troyer.
\newblock Progress towards practical quantum variational algorithms.
\newblock \emph{Physical Review A}, 92\penalty0 (4):\penalty0 042303, 2015.

\bibitem[Wiebe et~al.(2014)Wiebe, Granade, Ferrie, and Cory]{Wiebe_2014}
N.~Wiebe, C.~Granade, C.~Ferrie, and D.~Cory.
\newblock Hamiltonian learning and certification using quantum resources.
\newblock \emph{Physical Review Letters}, 112\penalty0 (19), 2014.
\newblock ISSN 1079-7114.
\newblock \doi{10.1103/physrevlett.112.190501}.
\newblock URL \url{http://dx.doi.org/10.1103/PhysRevLett.112.190501}.

\bibitem[Wiebe et~al.(2015)Wiebe, Granade, and Cory]{Wiebe_2015}
N.~Wiebe, C.~Granade, and D.~G. Cory.
\newblock Quantum bootstrapping via compressed quantum hamiltonian learning.
\newblock \emph{New Journal of Physics}, 17\penalty0 (2):\penalty0 022005, Feb. 2015.
\newblock ISSN 1367-2630.
\newblock \doi{10.1088/1367-2630/17/2/022005}.
\newblock URL \url{http://dx.doi.org/10.1088/1367-2630/17/2/022005}.

\bibitem[Wierstra et~al.(2011)Wierstra, Schaul, Glasmachers, Sun, and Schmidhuber]{wierstra2011natural}
D.~Wierstra, T.~Schaul, T.~Glasmachers, Y.~Sun, and J.~Schmidhuber.
\newblock Natural evolution strategies, 2011.

\bibitem[Wu and Koh(2023)]{wu2023errormitigated}
B.~Wu and D.~E. Koh.
\newblock Error-mitigated fermionic classical shadows on noisy quantum devices, 2023.

\bibitem[Yao et~al.(2020)Yao, Bukov, and Lin]{yao2020policy}
J.~Yao, M.~Bukov, and L.~Lin.
\newblock Policy gradient based quantum approximate optimization algorithm, 2020.

\bibitem[Yung et~al.(2014)Yung, Casanova, Mezzacapo, McClean, Lamata, Aspuru-Guzik, and Solano]{Yung2014From}
M.~H. Yung, J.~Casanova, A.~Mezzacapo, J.~McClean, L.~Lamata, A.~Aspuru-Guzik, and E.~Solano.
\newblock From transistor to trapped-ion computers for quantum chemistry.
\newblock \emph{Scientific Reports}, 4\penalty0 (1), 1 2014.
\newblock \doi{10.1038/srep03589}.

\bibitem[Zhang et~al.(2023)Zhang, Liu, Wu, Jiang, and Zhuang]{zhang2023dynamical}
B.~Zhang, J.~Liu, X.-C. Wu, L.~Jiang, and Q.~Zhuang.
\newblock Dynamical phase transition in quantum neural networks with large depth.
\newblock \emph{arXiv preprint arXiv:2311.18144}, 2023.

\bibitem[Zhao et~al.(2020)Zhao, Carleo, Stokes, and Veerapaneni]{Zhao_2020Natural}
T.~Zhao, G.~Carleo, J.~Stokes, and S.~Veerapaneni.
\newblock Natural evolution strategies and variational monte carlo.
\newblock \emph{Machine Learning: Science and Technology}, 2\penalty0 (2):\penalty0 02LT01, dec 2020.
\newblock \doi{10.1088/2632-2153/abcb50}.
\newblock URL \url{https://doi.org/10.1088%2F2632-2153%2Fabcb50}.

\end{thebibliography}

\end{document}